\documentclass[12pt]{article}
\pdfoutput=1
\usepackage{amsmath,amsfonts,graphicx,color,bbm,tikz}
\usepackage{comment}
\usepackage[nosort]{cite}
\usepackage{subfigure}
\usetikzlibrary{calc,positioning}
\usetikzlibrary{patterns,arrows,decorations.pathreplacing}
\usepackage{caption}
\usepackage{ulem}
\tikzset{>=stealth}

\makeatletter\@addtoreset{equation}{section}\makeatother

\newcommand{\be}{\begin{equation}}
\newcommand{\ee}{\end{equation}}
\newcommand{\bea}{\begin{eqnarray}}
\newcommand{\eea}{\end{eqnarray}}
\newcommand{\Tr}{{\rm Tr\,}}

\newcommand{\cS}{{\cal S}}

\newcommand{\bra}[1]{{\left< {#1} \right|}}
\newcommand{\ket}[1]{{\left| {#1} \right>}}

\def\nn{\nonumber}

\newcommand{\binomi}[2]{\begin{pmatrix} #1 \\ #2 \end{pmatrix}}

\renewcommand{\title}[1]{\vbox{\center\LARGE{#1}}\vspace{3mm}}
\renewcommand{\author}[1]{\vbox{\center#1}\vspace{3mm}}

\newcommand{\email}[1]{\vbox{\center\tt#1}\vspace{3mm}}

\begin{document}

\rightline{\small{\tt }}
\begin{center}

\vskip-1.5cm
{\large {\bf Area Law Violations and Quantum Phase Transitions in Modified Motzkin Walk Spin Chains} }
\vskip 0.75cm

 Fumihiko Sugino and Pramod Padmanabhan

\vskip 0.5cm 
Fields, Gravity \& Strings Group, Center for Theoretical Physics of the Universe,\\
Institute for Basic Science (IBS), Seoul 08826, Republic of Korea

\email{fusugino@gmail.com, pramod23phys@gmail.com}

\vskip 0.5cm 

\end{center}


\abstract{
\noindent 
Area law violations for entanglement entropy in the form of a square root have recently been studied for one-dimensional frustration-free quantum systems based on the Motzkin walks and their variations. Here we consider a Motzkin walk with a different Hilbert space on each step of the walk spanned by the elements of a {\it Symmetric Inverse Semigroup} with the direction of each step governed by its algebraic structure. This change alters the number of paths allowed in the Motzkin walk and introduces a ground state degeneracy that is sensitive to boundary perturbations.
We study the frustration-free spin chains based on three symmetric inverse semigroups, $\cS^3_1$, $\cS^3_2$ and $\cS^2_1$. 
The system based on $\cS^3_1$ and $\cS^3_2$ provide examples of quantum phase transitions in one dimension with the former exhibiting a transition between the area law and a logarithmic violation of the area law and the latter providing an example of transition from logarithmic scaling to a square root scaling in the system size, mimicking a colored $\cS^3_1$ system. The system with $\cS^2_1$ is much simpler and produces states that continue to obey the area law.

 }


\section{Introduction}

Local interacting Hamiltonians are good candidates for describing real systems and can be simulated on available computational resources in most cases. This is especially true for systems with low correlations as seen in the development of various techniques like density matrix renormalization groups (DMRG) and tensor networks \cite{dmrg, tens, whiteDMRG}, where the physically interesting degrees of freedom span a small subspace of the entire Hilbert space of the many body system, which is generally much larger. The entanglement entropy of the system can detect these correlations, and in particular gapped Hamiltonians in 1D possess ground states that obey the area law \cite{hastings, arad}, which is believed to hold in higher dimensions as well. Area laws for the entanglement entropy have also found uses in seemingly disparate areas of black hole physics, quantum information and quantum many body physics \cite{hetEE, eisertRev, 5i, ne, ne2}.

While we desire many body states obeying the area law to be able to simulate them, a generic quantum many body state is extensively entangled and obeys the volume law in subsystem size, seen in \cite{2i,3i,4i, 15i, 15s, 14i, 16i, 17i, bravyi}. Other examples include a supersymmetric many body system \cite{huijse}. For gapless systems we see a violation through a logarithmic factor \cite{16s}. Examples are found in $(1+1)$-dimensional conformal field theories \cite{7s, 18s, 10i, 9i} and in Fermi liquid theory \cite{19s, 13i}. 

More recently there have been constructions based on random walks, like the  {\it Motzkin} and {\it Dyck} walks (MWs or DWs respectively), that have led to yet another violation of the area law through a factor $\sqrt{n}$  \cite{shor, anna, dyck}. These models are frustration free, and have a unique ground state that is a uniform superposition of paths which satisfy the MWs or DWs. 
For the former case, the energy gap scales as a power law in the size of the system and the system is away from criticality 
(not described by conformal field theory). Such systems have been generalized by deforming the Hamiltonian with weights \cite{i1, kat2, i2} leading to a unique ground state that is now a weighted superposition of paths satisfying the MWs or DWs. For suitable values of the weights they show the area law and the volume law, thus modeling quantum phase transitions.  

In this paper we introduce a new modification of the MWs by restricting the walks with the elements of symmetric inverse semigroups. Inverse semigroups appear in a variety of areas that are of interest to physics. We convince the reader by providing some examples. They arise in the theory of partial symmetries and act as symmetries of the tilings of $\mathbb{R}^n$, and aperiodic structures like quasicrystals \cite{3p,6p,8p}. These structures are seldom used in quantum theory as they cannot be unitarily represented on Hilbert spaces due to its nature of only acting on subsets of the Hilbert space. Nevertheless their algebraic structure can be exploited to construct integrable supersymmetric many body systems \cite{PP}, which possess many-body localized states. 

\begin{figure}[h!]
\captionsetup{width=0.8\textwidth}
\begin{center}
		\includegraphics[scale=0.8]{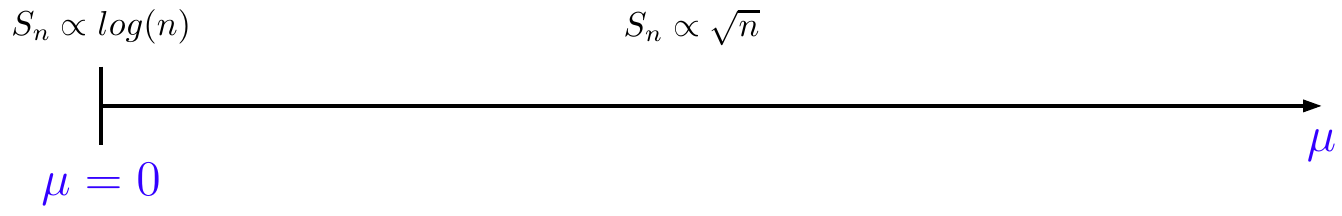} 
	\caption{\small The change in the behavior of the entanglement entropy as a function of a tunable parameter $\mu$ in the $\cS^3_2$ system of (\ref{S32Htotal_case2}), which is a colored version of the $\cS^3_1$ system .}
\label{phase1}
\end{center}
\end{figure}

Here we construct a frustration-free local Hamiltonian made up of local equivalence moves that are obtained using the structure of the inverse semigroups. The resulting systems model two kinds of quantum phase transitions measured by the entanglement entropy of the system, one between states obeying the area law and those violating it logarithmically in the system size, and another where the states either violate the area law logarithmically or violate it as the square root in the system size. These are summarized in Figs. \ref{phase} and \ref{phase1}. We also emphasize that the method used to compute the entanglement entropy is the standard one, as done in \cite{shor, anna, dyck}, which are for theories away from criticality, as we expect a similar behavior to hold for the models constructed here. Apart from this, these systems also possess a ground state degeneracy that arises due to the algebraic structure of the inverse semigroups. 

\begin{figure}[h!]
\captionsetup{width=0.8\textwidth}
\begin{center}
		\includegraphics[scale=0.8]{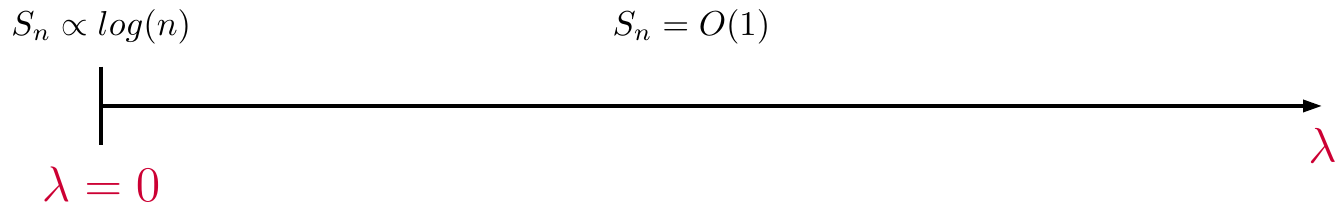} 
	\caption{\small The change in the behavior of the entanglement entropy as a function of a tunable parameter $\lambda$ in the $\cS^3_1$ system of (\ref{hs31t}).}
\label{phase}
\end{center}
\end{figure}

The paper is organized as follows. In Sec. \ref{2} we give the basic definition of inverse semigroups, recall the rules of the Motzkin walk and introduce its modification using the inverse semigroup elements. We build systems based on three symmetric inverse semigroups, namely $\cS^3_1$, $\cS^3_2$ and $\cS^2_1$. These are described in Secs. \ref{3}, \ref{sec:s32} and \ref{sec:s21} respectively. In each of these sections we present the frustration-free Hamiltonian built out of the local equivalence moves dictated by the algebraic structure of the inverse semigroup under study, find the number of paths that satisfy the modified Motzkin walk and compute the entanglement entropies of the half chain for the ground states. We conclude with a discussion and outlook in Sec. \ref{outlook}.

\section{MWs governed by Inverse Semigroups} \label{2}

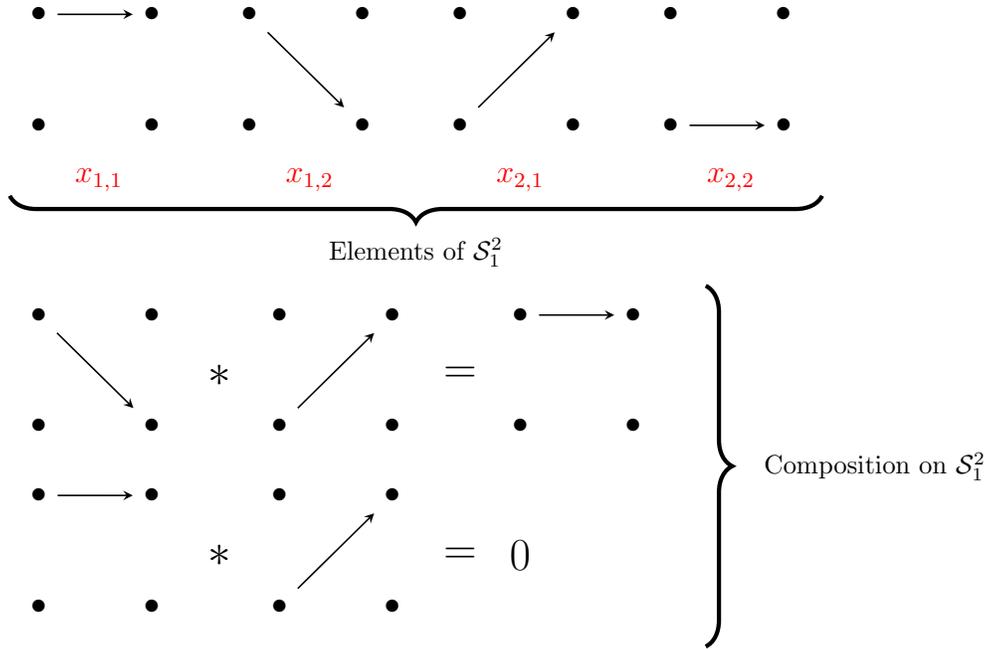
\begin{figure}[h!]
\captionsetup{width=0.8\textwidth}
\centering
\begin{tikzpicture}[scale=.8]
    \node (E) at (0,0) {$\bullet$};
    \node[right=of E] (F) {$\bullet$};
    \node[below=of F] (A) {$\bullet$};
    \node[below=of E] (B) {$\bullet$};
    \draw[->, semithick] (E)--(F) node at (1.0,-2.75) {$\textcolor{red}{x_{1,1}}$};
    
    \node (E2) at (3.5,0) {$\bullet$};
    \node[right=of E2] (F2) {$\bullet$};
    \node[below=of F2] (A2) {$\bullet$};
    \node[below=of E2] (B2) {$\bullet$};
    \draw[->,semithick] (E2)--(A2) node at (4.5,-2.75) {$\textcolor{red}{x_{1,2}}$};;
    
    \node (E3) at (7,0) {$\bullet$};
    \node[right=of E3] (F3) {$\bullet$};
    \node[below=of F3] (A3) {$\bullet$};
    \node[below=of E3] (B3) {$\bullet$};
    \draw[->,semithick] (B3)--(F3) node at (8,-2.75) {$\textcolor{red}{x_{2,1}}$};;
    
    \node (E4) at (10.5,0) {$\bullet$};
    \node[right=of E4] (F4) {$\bullet$};
    \node[below=of F4] (A4) {$\bullet$};
    \node[below=of E4] (B4) {$\bullet$};
    \draw[->,semithick] (B4)--(A4) node at (11.5,-2.75) {$\textcolor{red}{x_{2,2}}$};;
    
    \draw [ultra thick, decorate,decoration={brace,amplitude=10pt,mirror},xshift=0.5pt,yshift=-0.5pt](-0.5,-3) -- (13,-3) node[black,midway,yshift=-0.75cm] 
    {\footnotesize Elements of $\cS^2_1$};
    
    \node (E5) at (0,-5) {$\bullet$};
    \node[right=of E5] (F5) {$\bullet$};
    \node[below=of F5] (A5) {$\bullet$};
    \node[below=of E5] (B5) {$\bullet$};
    \draw[->,semithick] (E5)--(A5) node at (3,-6.0) {\Large $\ast$};;
    
    \node (E6) at (4,-5) {$\bullet$};
    \node[right=of E6] (F6) {$\bullet$};
    \node[below=of F6] (A6) {$\bullet$};
    \node[below=of E6] (B6) {$\bullet$};
    \draw[->,semithick] (B6)--(F6);; 
    
    \node (E7) at (8,-5) {$\bullet$};
    \node[right=of E7] (F7) {$\bullet$};
    \node[below=of F7] (A7) {$\bullet$};
    \node[below=of E7] (B7) {$\bullet$};
    \draw[->,semithick] (E7)--(F7) node at (7.0,-6) {\Large $=$};;
    
    \node (E8) at (0,-8) {$\bullet$};
    \node[right=of E8] (F8) {$\bullet$};
    \node[below=of F8] (A8) {$\bullet$};
    \node[below=of E8] (B8) {$\bullet$};
    \draw[->,semithick] (E8)--(F8) node at (3,-9.0) {\Large $\ast$};;
    
    \node (E9) at (4,-8) {$\bullet$};
    \node[right=of E9] (F9) {$\bullet$};
    \node[below=of F9] (A9) {$\bullet$};
    \node[below=of E9] (B9) {$\bullet$};
    \draw[->,semithick] (B9)--(F9)  node at (7,-9) {\Large $=$};;

    \node (E10) at (8.0,-9) {\Large $0$};
    \draw [ultra thick, decorate,decoration={brace,amplitude=10pt,mirror},xshift=130.4pt,yshift=-0.4pt](6.5,-10.5) -- (6.5,-4.5) node[black,midway,xshift=2.25cm] {\footnotesize Composition on $\cS^2_1$};
\end{tikzpicture}

	\caption{\small Diagrammatic representation of $\cS^2_1$. The composition rules are obtained by 
	tracing arrows connecting two elements. If one cannot trace a continuous arrow, the 
	product is 0.}
\label{s21}
\end{figure}

Semigroups can be thought of as generalizations of groups by relaxing the condition for the inverse. However we can still introduce a unique inverse to every element of the semigroup to make it an {\it inverse semigroup}. This is still different from a group, as now there is no unique identity element but instead partial identities. These ideas become clearer with {\it Symmetric Inverse Semigroups} (SISs) which are analogs of the permutation groups in group theory \cite{wagner-preston, mark}. Their elements are the partial functions on a set of finite order, $k$. It is denoted by $\cS^k_p$, where $p$ is the order of the subset on which the partial functions act. The partial symmetry elements of $\cS^k_1$ are denoted by $x_{a,b}$, with $a,b\in\{1,\cdots , k\}$, obeying the non-Abelian composition rule
\begin{equation}
 x_{a,b}\ast x_{c,d} = \delta_{bc}x_{a,d}.
 \label{SGproduct}
\end{equation}
The indices $a$ and $b$ can be thought of, respectively, as the domain and range of the partial symmetry operation. The elements and their algebra are best explained pictorially as in Fig. \ref{s21}. In this paper we work with three SISs, namely $\cS^3_1$, $\cS^3_2$ and $\cS^2_1$. 

To build a quantum system out of such SISs, we give it a Hilbert space structure as follows. First, we turn it into a vector space, spanned by the elements of $\cS^k_1$, $\{\ket{x_{a,b}}~;~a,b\in\{1,\cdots,k\}\}$ and use the canonical inner product, $\langle x_{a,b} \vert x_{c, d} \rangle = \delta_{ac} \delta_{bd}$. This is the equivalent of working in the regular representation of $\cS^k_1$. 
 
We can then consider random walks associated with $\cS^k_1$ on a two-dimensional  $(x,y)$ plane, by associating the above Hilbert space built using the elements of $\cS^k_1$ on each step and regarding $\ket{x_{a,b}}$ with $a<b$ as up moves, 
$a>b$ as down moves and $a=b$ as flat moves with the paths on $n$ steps starting at $(0,0)$ and ending at $(n,0)$ without entering the region $y<0$, as in the case of MWs or DWs. In the rest of this paper we will call these decorated walks as the {\it SIS Motzkin walks} (SMWs). 

With this modification of the MW, we see two features that distinguish the MWs and the SMWs. These are the kind of paths involved and the maximum heights reached. 

\paragraph{Kinds of paths} - 

The algebraic structure of $\cS^3_1$ introduces the idea of {\it connected}, {\it partially connected} and {\it disconnected} paths. The path is connected if two consecutive steps have elements of SISs that can be composed. Namely, $|x_{a,b}\rangle$ followed by $|x_{b,c}\rangle$ forms a connected path as the two can be composed by the algebra of $\cS^3_1$. If this condition is not satisfied for each step, it is disconnected and if only for some of the steps, it is partially connected. We remark on the dynamical implications later. 

\paragraph{Maximum heights} - 

\begin{figure}[h!]
\captionsetup{width=0.8\textwidth}
\begin{center}
		\includegraphics[scale=0.8]{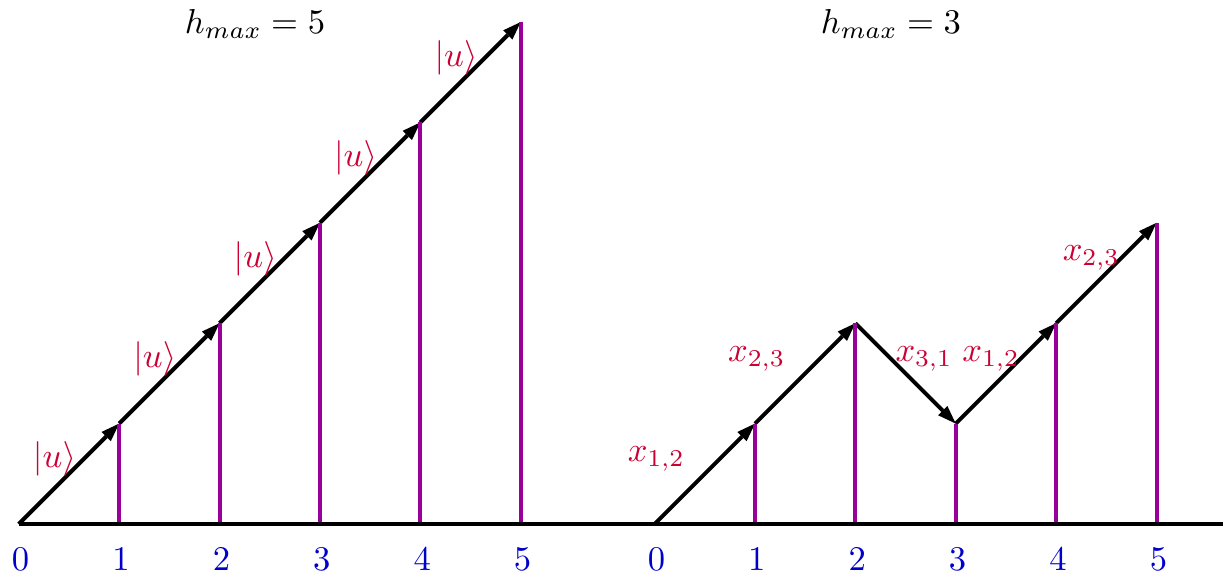} 
	\caption{\small The heights reached in the MW and the $\cS^3_1$ SMW for a 5-step walk.  
	The $\cS^3_1$ SMW is constrained by its algebraic structure. Note that by using other SISs $\cS^p_1$ with higher $p$ helps us reach far greater heights in a similar number of steps. }
\label{heights}
\end{center}
\end{figure}

Finally the maximum height we can reach in the $\cS^3_1$ SMWs is no longer $n$ as in the MWs of $2n$ steps but 
\begin{equation}
h_{max} = \left[\frac{n-2}{3}\right] + 2,
\end{equation}
with $\left[k\right]$ being the greatest integer not exceeding $k$, (see Fig. \ref{heights}). 

\section{The model based on $\cS^3_1$} \label{3}

\begin{figure}[h!]
\captionsetup{width=0.8\textwidth}
\begin{center}
		\includegraphics[scale=0.8]{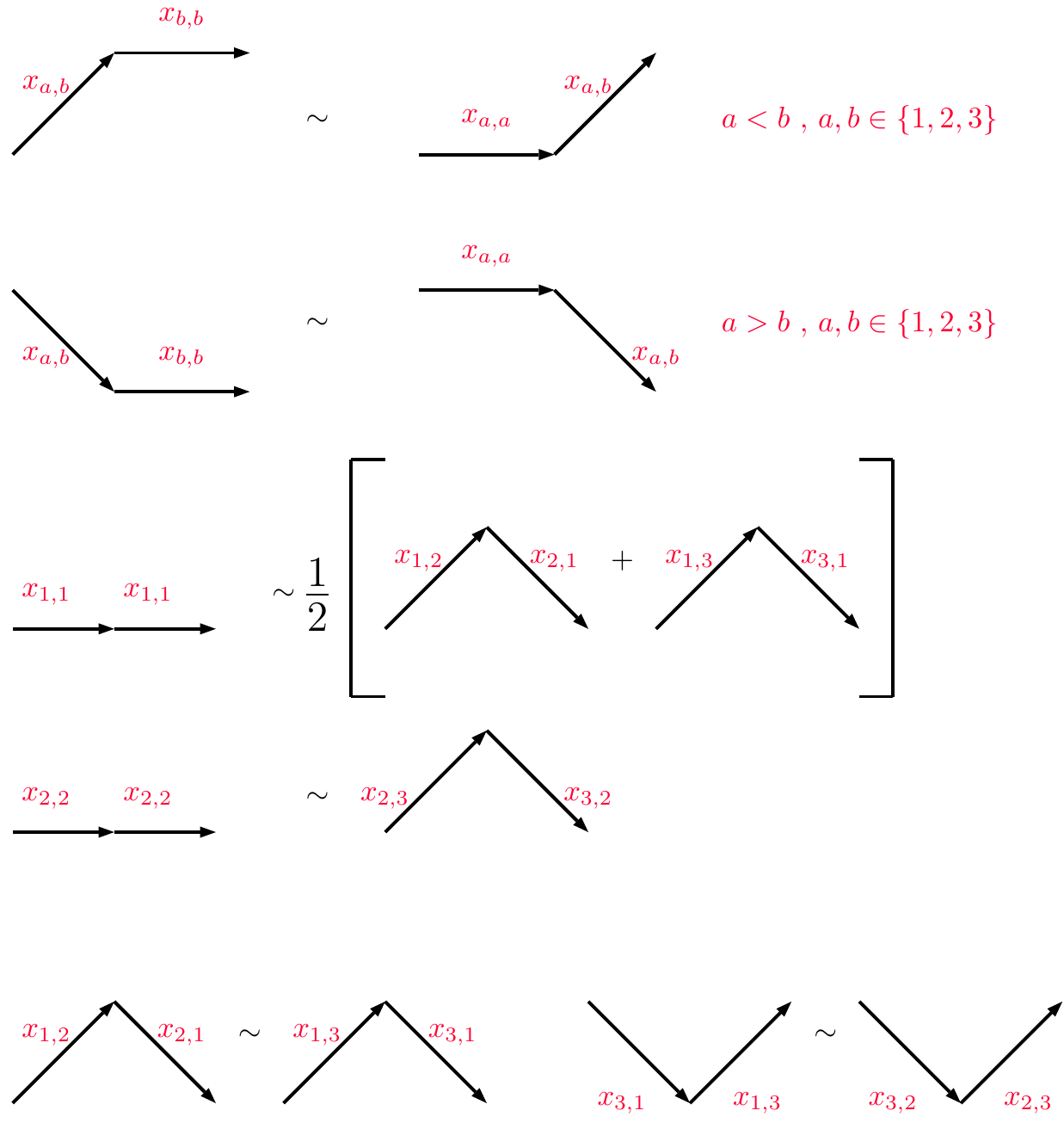} 
	\caption{\small The local equivalence moves for the $\cS^3_1$ SMWs. 
	The different ways of moving up and down do not amount to having different colors, as we shall see when we analyze this system.}
\label{leSM}
\end{center}
\end{figure}

The different paths satisfying the conditions of the $\cS^3_1$ SMWs can be mapped to each other using local equivalence moves illustrated in Fig. \ref{leSM}. To obtain the equal weight superposition of these paths as a ground state of a local, frustration-free Hamiltonian we need to project out these local moves using operators for the local equivalences for the ``up'' moves, ``down'' moves, ``flat'' moves and ``wedge'' moves given by

\begin{eqnarray}
\hspace{-7mm} U_{j, j+1}  & = & \sum_{a, b =1 ; a<b}^3~P^{\frac{1}{\sqrt{2}}\left[\ket{\left(x_{a,b}\right)_j, \left(x_{b,b}\right)_{j+1}} - \ket{\left(x_{a,a}\right)_j, \left(x_{a,b}\right)_{j+1}}\right]},  \\
\hspace{-7mm}D_{j, j+1}  & = & \sum_{a, b =1 ; a>b}^3~P^{\frac{1}{\sqrt{2}}\left[\ket{\left(x_{a,b}\right)_j, \left(x_{b,b}\right)_{j+1}} - \ket{\left(x_{a,a}\right)_j, \left(x_{a,b}\right)_{j+1}}\right]}, \\
\hspace{-7mm}F_{j, j+1} & = & P^{\sqrt{\frac23}\left[\ket{\left(x_{1,1}\right)_j, \left(x_{1,1}\right)_{j+1}} - \frac{1}{2}\left(\ket{\left(x_{1,2}\right)_j, \left(x_{2,1}\right)_{j+1}} + \ket{\left(x_{1,3}\right)_j, \left(x_{3,1}\right)_{j+1}}\right)\right]}  \nonumber \\
  &  & + P^{\frac{1}{\sqrt{2}}\left[\ket{\left(x_{2,2}\right)_j, \left(x_{2,2}\right)_{j+1}} - \ket{\left(x_{2,3}\right)_j, \left(x_{3,2}\right)_{j+1}}\right]} ,\\
\hspace{-7mm}W_{j, j+1} & = & P^{\frac{1}{\sqrt{2}}\left[\ket{\left(x_{1,2}\right)_j, \left(x_{2,1}\right)_{j+1}} - \ket{\left(x_{1,3}\right)_j, \left(x_{3,1}\right)_{j+1}}\right]} \nn \\
& & + \left(1-{\rm sgn}(\lambda)\right)  P^{\frac{1}{\sqrt{2}}\left[\ket{\left(x_{3,1}\right)_j, \left(x_{1,3}\right)_{j+1}} - \ket{\left(x_{3,2}\right)_j, \left(x_{2,3}\right)_{j+1}}\right]},
\end{eqnarray}
respectively, with $P^{\ket{\psi}}$ denoting a projector to the normalized state $\ket{\psi}$ 
and $\lambda (\geq 0)$ being a tunable parameter. 
These terms with $j\in\{1,\cdots, n-1\}$ make up the bulk Hamiltonian, $H_{bulk}$. 

The boundary terms prevent the walks from moving below the $x$-axis at the origin and upward at $(n,0)$ just as in the MW case. In the semigroup case this is taken care of on the left and right by
\begin{eqnarray}
H_{left}  & = & P^{\ket{\left(x_{2,1}\right)_1}} + P^{\ket{\left(x_{3,1}\right)_1}} + P^{ \ket{\left(x_{3,2}\right)_1}}, \\
H_{right}& = & P^{\ket{\left(x_{1,2}\right)_n}}  + P^{\ket{\left(x_{1,3}\right)_n}} + P^{\ket{\left(x_{2,3}\right)_n}}.
\end{eqnarray}

We can now also add a ``balancing'' term as in the MWs by including the term
\begin{equation}
B_{j, j+1} = P^{\ket{\left(x_{1,3}\right)_j, \left(x_{3,2}\right)_{j+1}}} + P^{\ket{\left(x_{2,3}\right)_j, \left(x_{3,1}\right)_{j+1}}}.
\end{equation} 
This term implies that if we go up with $\ket{\left(x_{1,3}\right)}$ or $\ket{\left(x_{2,3}\right)}$ we have to come down with $\ket{\left(x_{3,1}\right)}$ or $\ket{\left(x_{3,2}\right)}$ similar to what happens in the colored MW case, and is crucial for the phase transitions in this system. With this the total Hamiltonian is
\begin{equation} \label{hs31}
H = H_{left} + H_{bulk} +  H_{right} + \lambda\sum_{j=1}^{n-1}~B_{j, j+1}.
\end{equation}

This Hamiltonian has an extensive ground state degeneracy (GSD) due to the presence of partially connected paths and disconnected paths discussed in Sec. \ref{2}.
These paths can be lifted out of the ground states by adding the term 
\be
H_{bulk, \,disconnected} = \sum_{j=1}^{n-1}\sum_{a,b,c, d = 1; b\neq c}^3~ P^{\ket{\left(x_{a,b}\right)_j, \left(x_{c,d}\right)_{j+1}}}
\ee
to the Hamiltonian $H$ of (\ref{hs31}) making the total
\be \label{hs31t}
H_{\cS^3_1} = H + H_{bulk, \,disconnected}.
\ee

\subsection*{Reduced Hilbert Space and the construction of $H_{connected}$}
Here, we construct the Hamiltonian (called $H_{connected}$) describing the dynamics of just the connected paths 
(for not only the ground states but also excited states). 

\begin{figure}[h!]
\captionsetup{width=0.8\textwidth}
\begin{center}
		\includegraphics[scale=0.8]{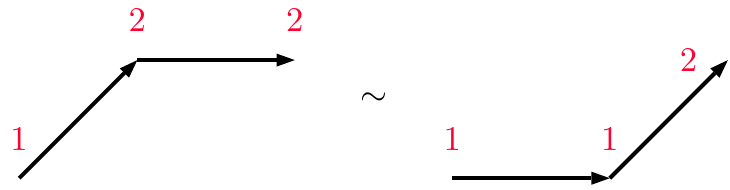} 
	\caption{\small The local equivalence moves for the $\cS^3_1$ SMWs in the reduced Hilbert space. For simplicity we show only one of the local equivalence moves.}
\label{leSMR}
\end{center}
\end{figure}

To capture the dynamics of just the {\it connected} paths we shift the local Hilbert space from the edges of the one-dimensional chain to the sites of the chain carrying a local Hilbert space of dimension 3, spanned by the states $\ket{1}, \ket{2}, \ket{3}$, which are just the semigroup indices. Now a path becomes a list of numbers which are naturally connected. For example on a 3-edge or 4-site chain we have $\ket{1,2,3,1} \equiv\ket{\left(x_{1,2}\right)_1, \left(x_{2,3}\right)_2, \left(x_{3,1}\right)_3}$. This change gives a new look to the local equivalence moves of Fig. \ref{leSM} as shown in Fig. \ref{leSMR}. 

The operators implementing these local equivalence moves and acting on the sites, $i\in \{1, \cdots, n-1\}$ in the bulk are now given by
\begin{eqnarray}
{\cal U}_i & = & \sum_{a,b=1 ; a<b}^3~P^{\ket{a}}_{i-1}~\frac12\left(1^{ab}_i-X^{ab}_i\right)~P^{\ket{b}}_{i+1} , \\
{\cal D}_i & = & \sum_{a,b=1 ; a>b}^3~P^{\ket{a}}_{i-1}~\frac12\left(1^{ab}_i-X^{ab}_i\right)~P^{\ket{b}}_{i+1}, 
\\
{\cal F}_i & = & P^{\ket{1}}_{i-1}~\frac13\left(1^{12}_i-X^{12}_i\right)~P^{\ket{1}}_{i+1} 
               +  P^{\ket{1}}_{i-1}~\frac13\left(1^{13}_i-X^{13}_i\right)~P^{\ket{1}}_{i+1} \nonumber \\
              & &  - P^{\ket{1}}_{i-1}~\frac16\left(1^{23}_i-X^{23}_i\right)~P^{\ket{1}}_{i+1}  
               +  P^{\ket{2}}_{i-1}~\frac12\left(1^{23}_i-X^{23}_i\right)~P^{\ket{2}}_{i+1}, \\
{\cal W}_i &  = & P^{\ket{1}}_{i-1}~\frac12\left(1^{23}_i-X^{23}_i\right)~P^{\ket{1}}_{i+1} 
                +  \left(1-{\rm sgn}(\lambda)\right) P^{\ket{3}}_{i-1}~\frac12\left(1^{12}_i-X^{12}_i\right)~P^{\ket{3}}_{i+1},  \\
 {\cal B}_i & = & P^{\ket{1}}_{i-1}~P^{\ket{3}}_i~P^{\ket{2}}_{i+1} + P^{\ket{2}}_{i-1}~P^{\ket{3}}_i~P^{\ket{1}}_{i+1},
\end{eqnarray}
with $1^{ab}$ and $X^{ab}$ being the partial identity and the partial $\sigma^x$ Pauli matrix in the $a$ and $b$ indices respectively:  
$1^{ab}=\ket{a}\bra{a}+ \ket{b}\bra{b}$ and $X^{ab}=\ket{a}\bra{b}+ \ket{b}\bra{a}$.
Each of these local operators acts on three sites as expected and constitutes $H_{bulk, connected}$ 
with a tunable parameter $\lambda$.

The boundary terms are now given by 
\begin{eqnarray}
H_{left, \,connected} & = & P^{\ket{2}}_0P^{\ket{1}}_1 + P^{\ket{3}}_0P^{\ket{1}}_1 + P^{\ket{3}}_0P^{\ket{2}}_1, \\ 
H_{right, \,connected} & = & P^{\ket{1}}_{n-1}P^{\ket{2}}_{n} + P^{\ket{1}}_{n-1}P^{\ket{3}}_{n} + P^{\ket{2}}_{n-1}P^{\ket{3}}_{n},
\end{eqnarray}
making the total {\it connected} Hamiltonian a sum of the boundary and bulk parts as before.

\subsection{Ground states } \label{4}

We discuss the structure of the ground states for the two cases of $\lambda=0$ and $\lambda> 0$ separately. 

\subsubsection{For $\lambda=0$ :} \label{l0}

The SIS algebra splits the paths according to different equivalence classes that cannot be mapped into each other by local equivalence moves of Fig. \ref{leSM}. This is seen by noting that we can start the walk with the semigroup index 1, 2 or 3. If we start with the semigroup index 1, that is with vectors $\ket{\left(x_{1,1}\right)}$, $\ket{\left(x_{1,2}\right)}$ or $\ket{\left(x_{1,3}\right)}$ on the first step, we can end with either 1 or 2 as the semigroup index on the last step. Thus, we get two equivalences classes denoted by $\{11\}$ and $\{12\}$. In a similar way we obtain the equivalence classes $\{22\}$, $\{21\}$ and $\{33\}$, (the last one being a product state), making the GSD 5, independent of the size of the chain. It is worth noting that this degeneracy does not arise due to the geometry or topology of the lattice but rather the algebraic structure of the SIS $\cS^3_1$. 

In particular, there is no symmetry transformation mapping one equivalence class to another 
(except reversing the paths exchanging $\{12\}$ and $\{21\}$), which will be evident by seeing 
that the number of the paths in each equivalence class is different as (\ref{Nij_asym}). 
The absence of the symmetry comes from the constraint of the MWs forbidding to enter the $y<0$ region. 
If we consider the states corresponding to random walks without the constraint, then we will have the 9 equivalence classes $\{ab\}$ 
($a,b=1,2,3$), which are mapped to each other by the permutation group $\cS^3$. 

This GSD is stable to local perturbations in the bulk of the chain that preserve the local equivalence moves, 
but is sensitive to the boundary perturbations that can lift some of the states out of the ground state sector. 
For example, a local perturbation at a boundary by $P^{\ket{1}}_0$ lifts $\{11\}$ and $\{12\}$ making the GSD 3. 

To understand the ground states in the different equivalence classes, we need to count the number of paths that satisfy
 the condition of the SMWs which is the normalization of these states. 

 \subsubsection*{Normalization of the ground states at $\lambda=0$}
 
\begin{figure}[h]
\captionsetup{width=0.8\textwidth}
\centering
\includegraphics[scale = 0.8]{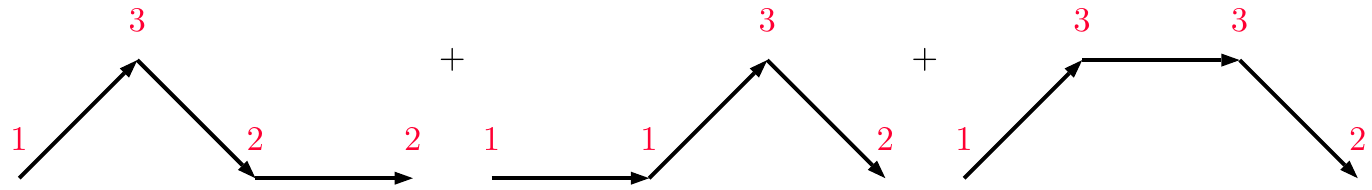}
\caption{\small Graphical expressions for the three terms of $P_{3,\, 1\to 2}$.} 
\label{fig:P3}
\end{figure}

 $P_{n,\,a\to b}$ denotes the formal sum of all possible connected paths on $n$ steps starting (ending) at the semigroup index $a$ ($b$). 
For example, $P_{3,\, 1\to 2}$ is given as 
\be P_{3,\, 1\to 2}= x_{1,3}x_{3,2}x_{2,2} + x_{1,1}x_{1,3}x_{3,2} + x_{1,3}x_{3,3}x_{3,2},\ee
shown in Fig.\ref{fig:P3}. 
Here and in what follows, the semigroup product ($\ast$) in (\ref{SGproduct}) is implicitly assumed for notational simplicity~\footnote{
Note that any path in the equivalence class $\{ab\}$ eventually reduces to $x_{a,b}$ as a consequence of the product (\ref{SGproduct}).
}. 
Let $N_{n,\,a\to b}$ be the number of walks included in $P_{n,\, a\to b}$, 
which is obtained by setting all the $x_{a,b}$ in $P_{n,\, a\to b}$ to 1; for instance, $N_{3,\, 1\to 2}=3$. 

We can see that 
$N_{n,\, a\to b}=N_{n,\, b\to a}$
by considering the reversed path starting from $(n,0)$ and ending at $(0,0)$. Also, $N_{n,\, 3\to a}=N_{n,\,a\to 3}=0$ with $a=1,2$, and $N_{n,\, 3 \to 3}=1$
as $P_{n,\, 3\to 3}= (x_{3,3})^n$. We use recursion relations to compute $N_{n, \, a\to b}$. 

\subsubsection*{Recursions for paths ending at height zero}
\begin{figure}[h]
\captionsetup{width=1\textwidth}
\centering
\includegraphics[scale = 0.8]{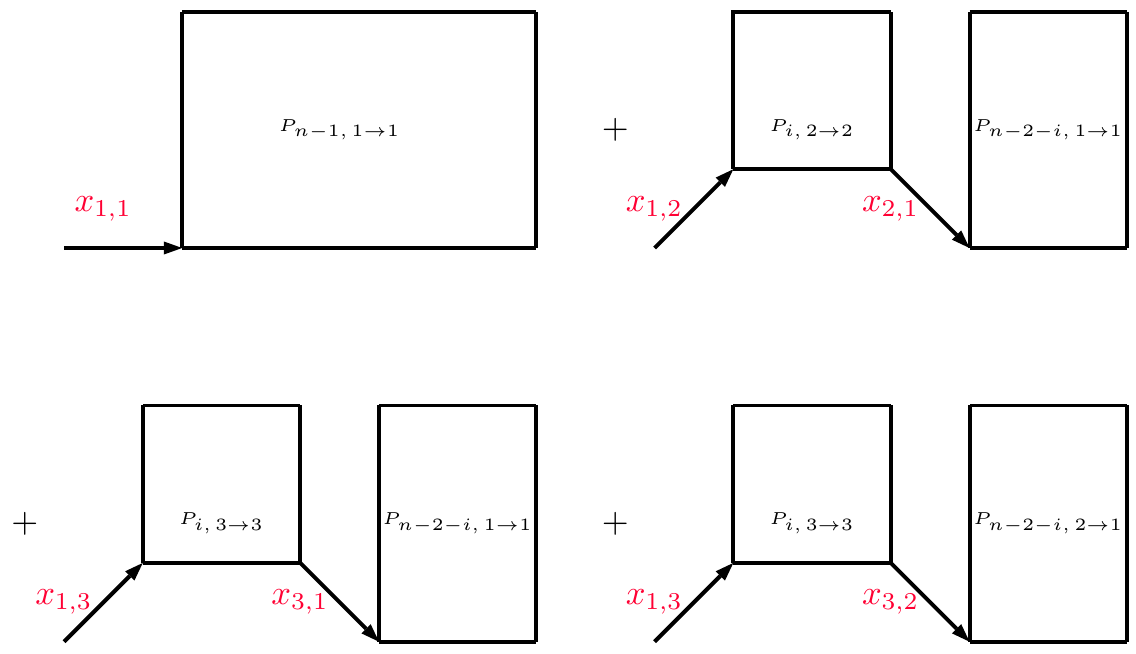}
\caption{\small The recursion in (\ref{P11_rec}) illustrated. } 
\label{recursion11}
\end{figure}

By looking at the first step of the walks, we can write down the following recursions (also see Fig.~\ref{recursion11}):
\bea
P_{n,\, 1\to 1} & = & x_{1,1}\,P_{n-1,\,1\to 1} +x_{1,2}\sum_{i=0}^{n-2}P_{i,\,2 \to 2}\, x_{2,1}\,P_{n-2-i,\,1\to 1} \nn \\
& &   +x_{1,3}\sum_{i=0}^{n-2}P_{i,\,3 \to 3}\, x_{3,1}\,P_{n-2-i,\,1\to 1} 
+x_{1,3}\sum_{i=0}^{n-2}P_{i,\,3 \to 3}\, x_{3,2}\,P_{n-2-i,\,2\to 1} , 
\label{P11_rec}
\\
P_{n,\, 2\to 2} & = & x_{2,2}\,P_{n-1,\,2\to 2} +x_{2,3}\sum_{i=0}^{n-2}P_{i,\,3 \to 3}\, x_{3,2}\,P_{n-2-i,\,2\to 2} 
+x_{2,3}\sum_{i=0}^{n-2}P_{i,\,3 \to 3}\, x_{3,1}\,P_{n-2-i,\,1\to 2}, 
\nn \\
\label{P22_rec}
\\
P_{n,\, 2\to 1} & = & x_{2,2}\,P_{n-1,\,2\to 1} +x_{2,3}\sum_{i=0}^{n-2}P_{i,\,3 \to 3}\, x_{3,2}\,P_{n-2-i,\,2\to 1} 
+x_{2,3}\sum_{i=0}^{n-2}P_{i,\,3 \to 3}\, x_{3,1}\,P_{n-2-i,\,1\to 1}. 
\nn \\
\label{P21_rec}
\eea

These lead to recursions for $N_{n,\, a\to b}$ as~\footnote{These are valid for $n\geq 1$. The terms of $\sum_{i=0}^{n-2}$ are regarded as null for $n=1$. } 
\bea
N_{n,\, 1\to 1} & = & N_{n-1,\,1\to 1} +\sum_{i=0}^{n-2}N_{i,\,2 \to 2}\, N_{n-2-i,\,1\to 1} \nn \\
& &   +\sum_{i=0}^{n-2}N_{i,\,3 \to 3}\, N_{n-2-i,\,1\to 1} 
+\sum_{i=0}^{n-2}N_{i,\,3 \to 3}\, N_{n-2-i,\,2\to 1} , 
\label{N11_rec}
\\
N_{n,\, 2\to 2} & = & N_{n-1,\,2\to 2} +\sum_{i=0}^{n-2}N_{i,\,3 \to 3}\, N_{n-2-i,\,2\to 2}  
+\sum_{i=0}^{n-2}N_{i,\,3 \to 3}\, N_{n-2-i,\,2\to 1}, 
\label{N22_rec}
\\
N_{n,\, 2\to 1} & = & N_{n-1,\,2\to 1} +\sum_{i=0}^{n-2}N_{i,\,3 \to 3}\, N_{n-2-i,\,2\to 1} 
+\sum_{i=0}^{n-2}N_{i,\,3 \to 3}\, N_{n-2-i,\,1\to 1}, 
\label{N21_rec}
\eea
where we use the invariance under the reversal property of the paths. By introducing the generating functions
\be
N_{a\to b}(x)\equiv \sum_{n=0}^\infty N_{n,\, a\to b}\, x^n \qquad \mbox{with} \qquad N_{0,\, a\to b}= \delta_{a, b},
\label{Nijg}
\ee
(\ref{N11_rec})-(\ref{N21_rec}) are recast as 
\bea
N_{1\to 1}(x) -1 & = & xN_{1\to 1}(x) +x^2N_{2\to 2}(x)N_{1\to 1}(x) +x^2N_{3\to 3}(x) N_{1\to 1}(x) \nn \\
& & +x^2 N_{3\to 3}(x) N_{2\to 1}(x) , 
\label{N11g_rec}
\\
N_{2\to 2}(x) -1 & = & xN_{2\to 2}(x) +x^2N_{3\to 3}(x) N_{2\to 2}(x) +x^2 N_{3\to 3}(x) N_{2\to 1}(x) , 
\label{N22g_rec}
\\
N_{2\to 1}(x) & = & xN_{2\to 1}(x) +x^2N_{3\to 3}(x) N_{2\to 1}(x) +x^2 N_{3\to 3}(x) N_{1\to 1}(x) . 
\label{N21g_rec}
\eea
Together with 
\be
N_{3\to 3}(x) =\frac{1}{1-x}
\label{N33g_sol}
\ee
which comes from $N_{n,\, 3 \to 3}=1$, we find that the equations become closed, and solve as 
\bea
N_{1\to 1}(x) & = & \frac{1-2x}{2x^3X}\left[1-\sqrt{1-4X^2}\right], 
\label{N11g_sol}
\\
N_{2\to 2}(x) & = & \frac{1-x}{2x^2(1-2x)}\left[1-2x- (1-2x-2x^2)\sqrt{1-4X^2}\right], 
\label{N22g_sol}
\\
N_{2\to 1}(x) & = & \frac{1}{2xX}\left[1-\sqrt{1-4X^2}\right]
\label{N21g_sol}
\eea
with 
\be 
X\equiv \frac{x^3}{(1-x)(1-2x-2x^2)}. 
\label{X}
\ee

Among the singularities of (\ref{N11g_sol}), (\ref{N22g_sol}) and (\ref{N21g_sol}), 
the nearest from the origin is $x=1/3$. Around this point, they behave as 
\bea
N_{1\to 1}(x) & = & 9-27\sqrt{3}\sqrt{1-3x} + O(1-3x), \nn \\
N_{2\to 2}(x) & = & 3-3\sqrt{3}\sqrt{1-3x} + O(1-3x), \nn \\
N_{2\to 1}(x) & = & 3-9\sqrt{3}\sqrt{1-3x}+ O(1-3x). 
\label{Nijg_sing}
\eea
By looking at the large order behavior in the expansion of $\sqrt{1-3x}$ around $x=0$, 
we can read off the large order behavior of the coefficients:
\be
N_{n,\, 1\to 1}  \sim  \frac{27\sqrt{3}}{2\sqrt{\pi}}\frac{3^n}{n^{3/2}}, \qquad 
N_{n, \,2\to 2}   \sim  \frac{3\sqrt{3}}{2\sqrt{\pi}}\frac{3^n}{n^{3/2}}, \qquad 
N_{n, \,2\to 1}   \sim  \frac{9\sqrt{3}}{2\sqrt{\pi}}\frac{3^n}{n^{3/2}} 
\label{Nij_asym}
\ee
as $n\to \infty$. 

\subsubsection*{Recursions for paths ending at nonzero height}
\begin{figure}[h]
\captionsetup{width=1\textwidth}
\centering
\includegraphics[scale = 0.8]{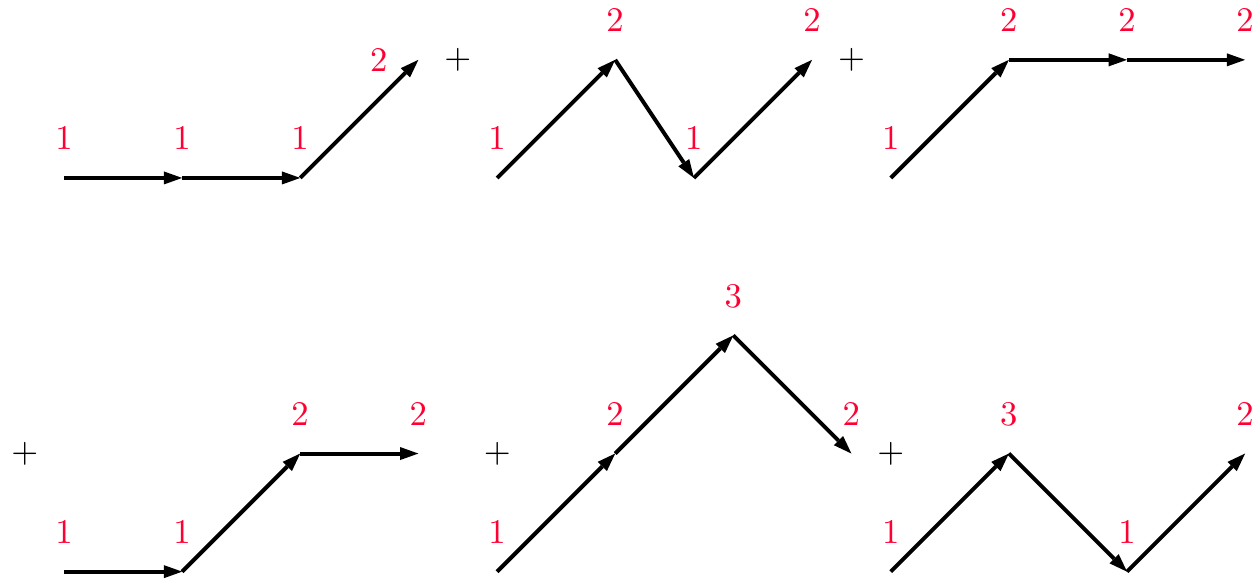}
\caption{\small Graphical expressions for the six terms of $P^{(1)}_{3,\, 1\to 2}$. } 
\label{fig:P(1)3}
\end{figure}

For later convenience, we also consider $n$-step walks obeying similar rules but starting at $(0,0)$ with the semigroup index $a$ and ending at $(n,h)$ with the index $b$. 
$h$ is a positive integer, and the paths never pass below the $x$-axis. 
$P^{(h)}_{n,\,a\to b}$ denotes the sum of such walks, and $N^{(h)}_{n,\, a\to b}$ counts the number of the walks in $P^{(h)}_{n,\, a\to b}$. 
For example, 
\bea
P^{(1)}_{3,\,1\to 2} & = & x_{1,1}x_{1,1}x_{1,2} + x_{1,2}x_{2,1}x_{1,2} + x_{1,2}x_{2,2}x_{2,2}  \nn \\
& & + x_{1,1} x_{1,2} x_{2,2} + x_{1,2}x_{2,3}x_{3,2} + x_{1,3}x_{3,1}x_{1,2}, \label{P(1)3}\\
N^{(1)}_{3,\, 1\to 2} & = & 6.
\eea 
The six paths in the r.~h.~s. of (\ref{P(1)3}) are depicted in Fig.~\ref{fig:P(1)3}. 
It is easy to see 
\be
N^{(h)}_{n,\, 3\to b}=0 \qquad \mbox{for} \quad b=1,2,3, \quad \mbox{and} \quad h\geq 1. 
\ee
Namely, there exists no path starting with the semigroup index 3 for any positive height. 

\begin{figure}[h]
\captionsetup{width=1\textwidth}
\centering
\includegraphics[scale = 0.8]{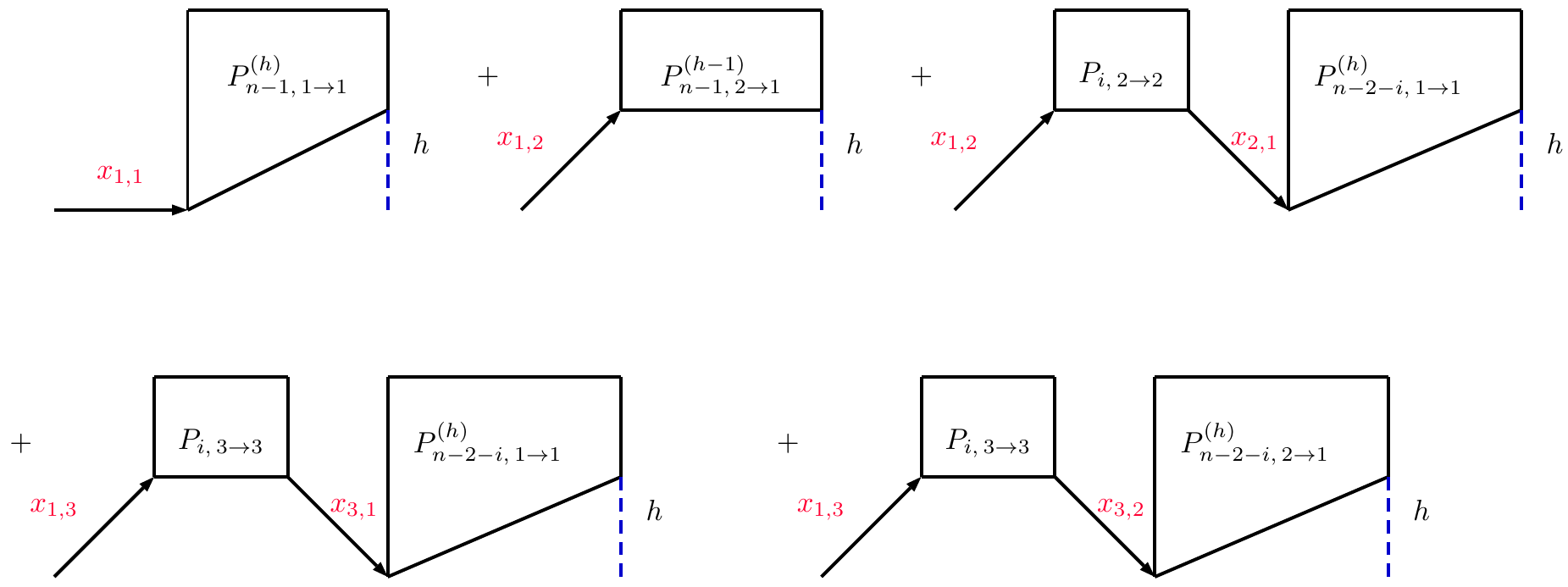}
\caption{\small The recursion in (\ref{Nh1j_rec}) illustrated for the $b=1$ case. } 
\label{recursion11h}
\end{figure}

We obtain recursion relations for these walks similar to the case of zero height. This is illustrated in Fig. \ref{recursion11h} for the $N^{(h)}_{n,\, 1\to 1}$ case.
The result is~\footnote{
Here, $N^{(0)}_{n,\, a\to b}$ is regarded as $N_{n,\, a\to b}$.
}   
\bea
N^{(h)}_{n,\, 1\to b} & = & N^{(h)}_{n-1,\, 1\to b} + N^{(h-1)}_{n-1,\, 2\to b} + \delta_{b, 3}\delta_{h,1} +\sum_{i=0}^{n-2}N_{i,\, 2\to 2}\, N^{(h)}_{n-2-i,\,1\to b} \nn \\
& & +\sum_{i=0}^{n-2}N_{i,\, 3\to 3}\, N^{(h)}_{n-2-i,\,1\to b} + \sum_{i=0}^{n-2}N_{i,\, 3\to 3}\, N^{(h)}_{n-2-i,\,2\to b} , 
\label{Nh1j_rec}
\\
N^{(h)}_{n,\, 2\to b} & = & N^{(h)}_{n-1,\, 2\to b} + \delta_{b, 3}\delta_{h,1} +\sum_{i=0}^{n-2}N_{i,\, 3\to 3}\, N^{(h)}_{n-2-i,\,1\to b}  + \sum_{i=0}^{n-2}N_{i,\, 3\to 3}\, N^{(h)}_{n-2-i,\,2\to b} , 
\nn \\
\label{Nh2j_rec}
\eea
for $b=1,2,3$, $n\geq 1$ and $h\geq 1$. 
In terms of the generating functions 
\be
N^{(h)}_{a\to b}(x)\equiv \sum_{n=0}^\infty N^{(h)}_{n,\, a\to b}\, x^n \qquad \mbox{with} \qquad N^{(h)}_{0,\,a\to b}=0
\label{Nhijg}
\ee
together with (\ref{Nijg}), we find that the pair of the equations (\ref{Nh1j_rec}) and (\ref{Nh2j_rec}) is closed for each $b$, and obtain 
\bea
N^{(h)}_{2\to 1}(x) & = & \frac{1}{x}\left(\frac{x^3}{1-2x} \,N_{1\to 1}(x)\right)^{h+1}, 
\label{Nh21g_sol}
\\
N^{(h)}_{1\to 1}(x) & = & \frac{1-2x}{x^2}\, N^{(h)}_{2\to 1}(x) = \frac{1-2x}{x^3}\left(\frac{x^3}{1-2x} \,N_{1\to 1}(x)\right)^{h+1}, 
\label{Nh11g_sol}
\\
N^{(h)}_{2\to 2}(x) & = & \left(\frac{x^3}{1-2x} \,N_{1\to 1}(x)\right)^{h}N_{2\to 2}(x), 
\label{Nh22g_sol}
\\
N^{(h)}_{1\to 2}(x) & = & \frac{1-2x}{x^2}\, N^{(h)}_{2\to 2}(x) = \frac{1-2x}{x^2}\left(\frac{x^3}{1-2x} \,N_{1\to 1}(x)\right)^{h}N_{2\to 2}(x), 
\label{Nh12g_sol}
\eea
\bea
N^{(h)}_{2\to 3}(x) & = & \frac{1}{x^2}\left(\frac{x^3}{1-2x} \,N_{1\to 1}(x)\right)^{h}\left[1-x-x^2N_{2\to 2}(x)\right],
\label{Nh23g_sol}
\\
N^{(h)}_{1\to 3}(x) & = & -\delta_{h,1}\frac{1}{x}+\frac{1-2x}{x^2}\,N^{(h)}_{2\to 3}(x) \nn \\
 & = & -\delta_{h,1}\frac{1}{x}+ \frac{1-2x}{x^4}\left(\frac{x^3}{1-2x} \,N_{1\to 1}(x)\right)^{h}\left[1-x-x^2N_{2\to 2}(x)\right]. 
\label{Nh13g_sol}
\eea
Plugging (\ref{Nijg_sing}) to these, we could read off the large order behavior of $N^{(h)}_{n,\, a\to b}$, which is useful for a fixed $h$ as $n\to \infty$ but not for cases for 
both of $n$ and $h$ growing. 
In order to find useful expressions even for the latter cases, we compute the number of the DWs by two different ways in appendix~\ref{app:Dyck}. 
As a result, we obtain the identity:
\be
X^h\left\{\frac{1}{2X^2}\left(1-\sqrt{1-4X^2}\right)\right\}^{h+1}=\sum_{n=0}^\infty N_n^{(h)}X^n , 
\label{Id}
\ee
where $N^{(h)}_n$ can be nontrivial only when $n$ and $h$ have the same parity (even/odd) and then takes 
\be
N^{(h)}_n=\frac{h+1}{\frac{n+h}{2}+1} \binomi{n}{\frac{n+h}{2}}.
\label{Nhn}
\ee

\subsubsection*{$\boldsymbol{N^{(h)}_{p,\, 2\to 1}}$ :} 
First, let us obtain the large order behavior of $N^{(h)}_{p,\, 2\to 1}$ as $p\to \infty$. 
Applying (\ref{Id}) to (\ref{Nh21g_sol}) with (\ref{N11g_sol}) leads to
\be
N^{(h)}_{2\to 1}(x)=\frac{1}{x}\sum_{n=0}^\infty N^{(h)}_nX^{n+1}. 
\ee
From (\ref{X}), $X^{n+1}$ has the expansion 
\be
X^{n+1}= \sum_{k,\ell=0}^\infty \sum_{j=0}^\ell \binomi{n+k}{k} \binomi{n+\ell}{\ell}\binomi{\ell}{j}2^\ell \,x^{3n+k+\ell+j+3}.
\ee
Plugging these two, we have 
\be
N^{(h)}_{p,\,2\to 1}=\sum_{n,\ell\geq 0}^*\sum_{j=0}^\ell N^{(h)}_n\binomi{p-2n-\ell-j-2}{n}\binomi{n+\ell}{\ell}\binomi{\ell}{j}2^\ell ,
\label{Nh21_sol}
\ee
where the asterisk (*) put to the first summation means $n$ and $\ell$ running under the condition $p-3n-\ell-j\geq 2$. 
It is found that the summand of (\ref{Nh21_sol}) has a saddle point (a stable point with respect to the deviations $n\to n+2$, $\ell\to \ell+1$ and $j\to j+1$) at
\be
n\sim\frac{2}{27}p, \qquad \ell\sim\frac{16}{27}p, \qquad j\sim\frac{4}{27}p 
\label{saddle}
\ee
for $p$ large. 

By using Stirling's formula ($n!\simeq \sqrt{2\pi}\,n^{n+\frac12}e^{-n}$), 
the asymptotic form of $N^{(h)}_n$ becomes 
\bea
N^{(h)}_n & \simeq & (h+1)\frac{2^{3/2}}{\sqrt{\pi}}\frac{2^n}{n^{3/2}}\,
\exp\left[1-\frac{n+h+3}{2}\ln\left(1+\frac{h+2}{n}\right)-\frac{n-h+1}{2}\ln\left(1-\frac{h}{n}\right)\right] \nn \\
& & \times \left[1+O\left(n^{-1}\right)\right]. 
\eea
The power of the exponential is expanded in large $n$ as 
\be
-\frac{(h+1)^2}{2n}-\frac{3}{2n}+\frac{(h+1)^2}{n^2} + \frac{2}{3n^2} +O\left(\frac{h^4}{n^3}\right), 
\ee
in which the first term provides the Gaussian factor rapidly decaying for $h> \sqrt{n}$. 
Bringing down the other terms from the exponential, we obtain  
\be
N^{(h)}_n\simeq (h+1)\frac{2^{3/2}}{\sqrt{\pi}}\frac{2^n}{n^{3/2}}\, e^{-\frac{1}{2n}(h+1)^2} 
\times \left[1+\frac{(h+1)^2}{n^2} + O\left(\frac{1}{n},\,\frac{h^4}{n^3}\right) \right]. 
\label{Nhn_asym0}
\ee
Note that due to the Gaussian factor, the order of $h$ is effectively at most $O\left(\sqrt{n}\right)$. So, we can regard 
the terms of $\frac{(h+1)^2}{n^2}$ and $\frac{h^4}{n^3}$ as $O\left(n^{-1}\right)$ quantities. 
(\ref{Nhn_asym0}) can be written as 
\be
N^{(h)}_n\simeq (h+1)e^{-\frac{1}{2n}(h+1)^2} \, N^{(0)}_n \times \left[1+O\left(n^{-1}\right)\right]. 
\label{Nhn_asym}
\ee 

We plug (\ref{Nhn_asym}) to (\ref{Nh21_sol}) and replace the exponential factor $e^{-\frac{1}{2n}(h+1)^2}$ with that at the saddle point (\ref{saddle}). The fluctuation around the saddle point in the sum over $n$ can be neglected for large $p$ as seen in appendix~\ref{app:fluctuation}. 
Then, $h$-dependence of (\ref{Nh21_sol}) can be pulled out of the sum to yield 
\bea
N^{(h)}_{p,\, 2\to 1} & \simeq &  (h+1)e^{-\frac{27}{4p}(h+1)^2} N_{p,\,2\to 1}\times \left[1+O\left(p^{-1}\right)\right] \nn \\
& \sim & (h+1)e^{-\frac{27}{4p}(h+1)^2} \, \frac{9\sqrt{3}}{2\sqrt{\pi}}\frac{3^p}{p^{3/2}} \times \left[1+O\left(p^{-1}\right)\right],
\label{Nh21_asym}
\eea
where we used (\ref{Nij_asym}) at the last step. 

\subsubsection*{Other coefficients:}
Once we know (\ref{Nh21_asym}), it is straightforward to obtain the large order behavior for the other coefficients from (\ref{Nh21g_sol})-(\ref{Nh13g_sol}). 
For instance, we find $N^{(h)}_{p,\,1\to 1}= N^{(h)}_{p+2,\,2\to 1}-2N^{(h)}_{p+1,\,2\to 1}$ from (\ref{Nh11g_sol}). 
Eventually, we have the expressions (up to multiplicative factors of $\left[1+O\left(p^{-1}\right)\right]$):
\bea
N^{(h)}_{p,\,1\to 1} 
& \sim & (h+1) e^{-\frac{27}{4p}(h+1)^2} \, \frac{27\sqrt{3}}{2\sqrt{\pi}}\frac{3^p}{p^{3/2}},
\label{Nh11_asym}
\\
N^{(h)}_{p,\,1\to 2}  
& \sim & \left[2h e^{-\frac{27}{4p}h^2}+(h+1) e^{-\frac{27}{4p}(h+1)^2} \right]\frac{9\sqrt{3}}{2\sqrt{\pi}}\frac{3^p}{p^{3/2}},
\label{Nh12_asym}
\\
N^{(h)}_{p,\, 2\to 2}&\sim & \left[2h e^{-\frac{27}{4p}h^2}+(h+1) e^{-\frac{27}{4p}(h+1)^2} \right]\frac{3\sqrt{3}}{2\sqrt{\pi}}\frac{3^p}{p^{3/2}},
\label{Nh22_asym}
\\
N^{(h)}_{p,\,2\to 3}  
& \sim & \left[4h e^{-\frac{27}{4p}h^2}-(h+1) e^{-\frac{27}{4p}(h+1)^2} \right]\frac{3\sqrt{3}}{2\sqrt{\pi}}\frac{3^p}{p^{3/2}},
\label{Nh23_asym}
\\
N^{(h)}_{p,\,1\to 3}  
& \sim & \left[4h e^{-\frac{27}{4p}h^2}-(h+1) e^{-\frac{27}{4p}(h+1)^2} \right]\frac{9\sqrt{3}}{2\sqrt{\pi}}\frac{3^p}{p^{3/2}}. 
\label{Nh13_asym}
\eea
 
 As a consistency check, we can see that (\ref{Nh21_asym})-(\ref{Nh13_asym}) together with (\ref{Nij_asym}) satisfy the composition law~\footnote{
Note that the number of $p$-step paths from the height $h$ with the semigroup index $b$ to the height 0 with the index $c$ is equal to $N^{(h)}_{p, \,c\to b}$. 
The sum over $h$ can be computed by converting it to the integral. 
}:
 \be
 \sum_{h=0}^\infty\sum_{b=1}^3 N^{(h)}_{p,\,a\to b} N^{(h)}_{p,\,c\to b} =N_{2p,\, a\to c},
 \label{composition}
 \ee
except for errors of $O\left(p^{-1}\right)$. 

\subsubsection{For $\lambda> 0$ : }

In this case there is a drastic change in the behavior of the ground states as the maximum height we can reach in a given path is only two. This is due to the fact that we are no longer allowed to come down by $\ket{\left(x_{3,1}\right)}$ (or $\ket{\left(x_{3, 2}\right)}$) once we go up by $\ket{\left(x_{2,3}\right)}$ (or $\ket{\left(x_{1,3}\right)}$). Thus we lose two of the equivalence classes, $\{12\}$ and $\{21\}$, reducing the GSD from 5 to 3. 

Switching on $\lambda$ changes the recursion relations of the generating functions for computing the normalization of the state in the zero height case to 
\begin{eqnarray}
N_{1\to 1}(x) - 1 & = & xN_{1\to 1}(x) + x^2\left(N_{2\to 2}(x)N_{1\to 1}(x) + N_{3\to 3}(x)N_{1\to 1}(x)\right), \\
N_{2\to 2}(x) - 1 & = & xN_{2\to 2}(x) + x^2N_{3\to 3}(x)N_{2\to 2}(x), 
\end{eqnarray}
with $N_{3\to 3}(x) = \frac{1}{1-x}$ as in the $\lambda = 0$ case. 

Solving these we obtain
\begin{eqnarray}
N_{1\to 1}(x) & =  & \frac{(x-1)(1-2x)}{1-4x+3x^2+2x^3-x^4}, \\
N_{2\to 2}(x) & = & \frac{1-x}{1-2x}.
\end{eqnarray}

The leading order behavior of the coefficients of these two terms is given by
\begin{equation}
N_{1\to 1} \sim \frac{(3+\sqrt{5})^n(\sqrt{5}+1)}{ 2^{n+2}\sqrt{5}}, ~~ N_{2\to 2} \sim 2^{n-1}.
\end{equation}

The generating functions for the normalizations in the case of nonzero heights is given by
\begin{eqnarray}
N^{(1)}_{1\to 2}(x) & = & \frac{x(1-x)^2}{1-4x+3x^2+2x^3-x^4}, \\ 
N^{(1)}_{1\to 3}(x) & = & \frac{x(1-2x)}{1-4x+3x^2+2x^3-x^4}, \\ 
N^{(2)}_{1\to 3}(x) & = & \frac{x^2(1-x)}{1-4x+3x^2+2x^3-x^4}, \\ 
N^{(1)}_{2\to 3}(x) & = & \frac{x}{1-2x}.
\end{eqnarray}
The rest are zero due to the height restriction as noted earlier. 

The leading order behavior of the coefficients of the first three terms is given by
\begin{equation}
N^{(1)}_{1\to 2} \sim \frac{(3+\sqrt{5})^n(\sqrt{5}+1)}{ 2^{n+2}\sqrt{5}}, ~ N^{(1)}_{1\to 3} \sim \frac{(3+\sqrt{5})^n}{ 2^{n+1}\sqrt{5}}, ~ N^{(2)}_{1\to 3} \sim \frac{(3+\sqrt{5})^n}{ 2^{n+1}\sqrt{5}}.
\end{equation}

\subsection{Entanglement entropy of the ground states}
\label{sec:EEGS}

\subsubsection{For $\lambda = 0$ :}

The normalized ground state $\{11\}$ for the system of the length $2n$ is expressed as 
\be
\ket{P_{2n,\, 1\to 1}} =\frac{1}{\sqrt{N_{2n,\,1\to 1}}} \,\sum_{w\in P_{2n,\,1\to 1}} \ket{w},
\label{GS11}
\ee
where $w$ runs over paths in $P_{2n,\,1\to 1}$. We split the system of length $2n$ into two equal subsystems A and B. 
Consider paths in $P_{2n,\, 1\to 1}$ that reach the point $(n,h)$, with the semigroup index $a$. 
The paths belonging to A are $P^{(h)}_{n,\,1\to a} \equiv P^{(0\to h)}_{n,\,1 \to a}$ to denote that it starts at height 0 and ends at height $h$. And for B we have the reversed paths of 
$P^{(h)}_{n,\,1\to a}$ denoted by $P^{(h\to 0)}_{n,\, a\to 1}$. 
Their corresponding normalized states are expressed by 
\be
\ket{P^{(0\to h)}_{n,\,1\to a}} = \frac{1}{\sqrt{N^{(h)}_{n,\,1\to a}}} \,\sum_{w\in P^{(0\to h)}_{n,\,1\to a}} \ket{w},
\qquad 
\ket{P^{(h\to 0)}_{n,\, a\to 1}} = \frac{1}{\sqrt{N^{(h)}_{n,\,1\to a}}} \,\sum_{w\in P^{(h\to 0)}_{n,\,a\to 1}} \ket{w},
\label{GShi1}
\ee
respectively. 

The Schmidt decomposition of (\ref{GS11}) leads to the following formula for the entanglement entropy
\be
\ket{P_{2n,\, 1\to 1}} =\sum_{h\geq 0}\sum_{a=1}^3\sqrt{p^{(h)}_{n,\, 1\to a\to1}}\, \ket{P^{(0\to h)}_{n,\,1\to a}} \otimes \ket{P^{(h\to 0)}_{n,\, a\to 1}} , 
\label{GS11decomp}
\ee
where $p^{(h)}_{n,\, 1\to a\to1}\equiv \frac{\left(N^{(h)}_{n,\,1\to a}\right)^2}{N_{2n,\,1\to 1}}$ 
satisfies
\be
\sum_{h\geq 0}\sum_{a=1}^3 p^{(h)}_{n,\, 1\to a\to1}=1
\ee
 due to (\ref{composition}) with $a=c=1$.

The reduced density matrix for the subsystem A takes the diagonal form as 
\be
\rho_{A, \, 1\to 1} = \Tr_B \ket{P_{2n,\, 1\to 1}} \,\bra {P_{2n,\, 1\to 1}} 
=\sum_{h\geq 0}\sum_{a=1}^3 p^{(h)}_{n,\, 1\to a\to1} \ket{P^{(0\to h)}_{n,\,1\to a}} \, \bra {P^{(0\to h)}_{n,\,1\to a}}, 
\ee
from which the entanglement entropy reads 
\be
S_{A,\, 1\to 1}= -\sum_{a=1}^3\sum_{h\geq 0} p^{(h)}_{n,\, 1\to a\to1}\, \ln p^{(h)}_{n,\, 1\to a\to1}.
\label{EE11}
\ee

By using (\ref{Nij_asym}), (\ref{Nh11_asym}), (\ref{Nh12_asym}) and (\ref{Nh13_asym}),  
we find the logarithmic violation of the area law:
\be
S^{\lambda=0}_{A, \, 1\to 1}=\frac12\ln n + \frac12\ln\frac{2\pi}{3} +\gamma -\frac12 +(\mbox{terms vanishing as $n\to \infty$})
\label{EE11f}
\ee
with $\gamma$ being the Euler constant. 
This behavior including the constant term is exactly the same as the uncolored Motzkin spin chain ($s=1$)~\cite{shor} and the quantum spin-1 chain at criticality~\cite{bravyi}, which was proposed in \cite{kit8, kit9} and proved in \cite{kore}. For other ground states in equivalence classes $\{12\}$, $\{21\}$ and $\{22\}$, we obtain the same result except for $\{33\}$ whose entanglement entropy is zero.  

\subsubsection{For $\lambda> 0$ :}

The computation for the entanglement entropy follows the same steps as in the $\lambda = 0$ case. However, in this case we only have three ground states, $\{11\}$, $\{22\}$ and $\{33\}$, as compared to the five ground states of the $\lambda=0$ case. The ground state based on the equivalence class $\{33\}$ is still a product state as before and hence has zero entanglement entropy. 
We use (\ref{EE11}) to compute the entanglement entropy for the $\{11\}$ ground state using
\begin{eqnarray}
p^{(h)}_{n, \,1\to a\to1} & = & \frac{\sqrt{5}+1}{4\sqrt{5}}~;~a = 1, h=0, \mbox{ or }a= 2, h=1 \\
p^{(h)}_{n, \,1\to 3\to1} & = & \frac{1}{\sqrt{5}(\sqrt{5}+1)}~;~h = 1, 2,
\end{eqnarray}
which gives
\begin{equation}
S^{\lambda> 0}_{A, \, 1\to 1} = \frac{1}{2\sqrt{5}}\left[-(\sqrt{5}+1)\ln(\sqrt{5}+1) - (\sqrt{5}-1)\ln(\sqrt{5}-1) + 2\sqrt{5}\ln(4\sqrt{5})\right]
\end{equation}
independent of $\lambda$, implying the area law, which is a feature also seen in the deformed MW cases \cite{i3}. 
For the case when the ground state is based on the equivalence class $\{22\}$, a similar result holds
 ($S^{\lambda>0}_{A,\, 2\to 2}=\ln 2$). 

With this result we explain the quantum phase transition in the phase diagram in Fig. \ref{phase}. 

\section{The model based on $\cS^3_2$} \label{sec:s32}

The SIS, $\cS^3_2$ has 18 elements and are denoted by $x_{ab, \, cd}$ with $ab \in \{12, 23, 31\}$ and $cd\in \{12, 23, 31, 21, 32, 13\}$. 
They satisfy the algebraic relation 
\begin{equation}
x_{ab, \, cd} * x_{ef, \, gh} = \delta_{ce}\delta_{df}~ x_{ab, \, gh} + \delta_{cf}\delta_{de} ~x_{ab, \, hg}.
\end{equation} 
We can equally realize them in terms of two sets of nine elements each, which we denote by $E = \{e_{a,b};~a,b\in\{1,2,3\}\}$ and $Z = \{z_{a,b};~a,b\in\{1,2,3\}\}$ respectively. The elements of these two sets will have $\cS^3_1$ indices. These are given by
\begin{equation}
\begin{array}{ccc}
e_{1,1} = x_{12,\, 12}, \, & e_{2,1} = x_{23,\, 12},\, & e_{3,1} = x_{31,\, 12}, \\
e_{1,2} = x_{12,\, 23}, \, & e_{2,2} = x_{23,\, 23}, \, & e_{3,2} = x_{31,\, 23}, \\ 
e_{1,3} = x_{12,\, 31}, \, & e_{2,3} = x_{23,\, 31}, \, & e_{3,3} = x_{31,\, 31},
\end{array}
\end{equation} 
and 
\begin{equation}
\begin{array}{ccc}
z_{1,1} = x_{12,\, 21}, \, & z_{2,1} = x_{23,\, 21},\, & z_{3,1} = x_{31,\, 21}, \\
z_{1,2} = x_{12,\, 32}, \, & z_{2,2} = x_{23,\, 32}, \, & z_{3,2} = x_{31,\, 32}, \\ 
z_{1,3} = x_{12,\, 13}, \, & z_{2,3} = x_{23,\, 13}, \, & z_{3,3} = x_{31,\, 13}.
\end{array}
\end{equation} 

It is easy to verify the algebra satisfied by the elements of $E$ and $Z$:
\begin{eqnarray}
e_{a,b}*e_{c,d} = \delta_{b,c}e_{a,d}, \\
 e_{a,b}*z_{c,d} = z_{a,b}*e_{c,d} = \delta_{b,c}z_{a,d}, \\ 
z_{a,b}*z_{c,d}  = \delta_{b,c}e_{a,d}.
\end{eqnarray}  
Thus $E$ and $Z$ satisfy the relations of $\mathbb{Z}_2$ as sets. This realization of the SIS $\cS^3_2$ in terms of the sets $E$ and $Z$ can be generalized with higher SISs, $\cS^n_p$ for $p>1$, being realized in terms of $p!$ sets $Z_s = \left[z^s_{a,b};~a,b\in\{1, \cdots, \binomi{n}{p}\}\right]$ with $s\in\{1, \cdots, p!\}$. The $p!$ sets, $Z_s$ satisfy the relations of the permutation group $S_p$. We will not prove this or dwell on the higher SISs in this paper. 

With this realization we can think of the SIS $\cS^3_2$ as a {\it colored} $\cS^3_1$ SMW, with the {\it color} degrees of freedom denoted by $s = \{1,2\}$. 
With this interpretation we denote the elements of $\cS^3_2$ as $\left[x^s_{a,b};~ a, b\in\{1, 2, 3\}\right]$, 
identifying $x^1_{a, b} = e_{a, b}$ and $x^2_{a, b} = z_{a, b}$.

\begin{figure}[h!]
\captionsetup{width=0.8\textwidth}
\begin{center}
		\includegraphics[scale=1]{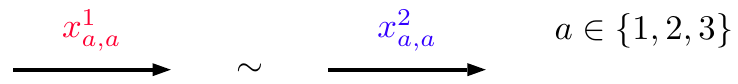} 
	\caption{\small The local equivalence move depicting the equivalence of the flat steps in the $\cS^3_2$ case. }
\label{leC}
\end{center}
\end{figure}

We can now write down the Hamiltonian for this case using the local equivalence moves shown in Fig. \ref{leSM}, 
with the change being that the $\cS^3_1$ elements on the links now get color degrees of freedom, $s \in \{1, 2\}$. 
In addition, we introduce the equivalence moves for the flat walks with only the colors changed: 
\be
x^1_{a,a}\sim x^2_{a,a} \qquad (a=1,2,3), 
\label{S32Cmove}
\ee
shown in Fig.~\ref{leC}. 

Here, we consider the following two cases regarding random walks realized by configurations in the ground states. 
In the first case, each step of the $\cS^3_1$ SMWs is independently duplicated by the color degrees of freedom. 
On the other hand, in the second case, the color degrees of freedom should be matched for each pair of up and subsequent down steps, 
but the rest is the same as the first case. 

\subsection{Hamiltonian for the first case}
The Hamiltonian for the first case is given by
\be
H^{colored}=\sum_{j=1}^n C_j +\sum_{j=1}^{n-1}\left[U_{j, j+1} + D_{j, j+1} + F^{colored}_{j, j+1}+ W^{colored}_{j, j+1} \right]
+ H_{left} + H_{right}
\label{S32H1st}
\ee
with the bulk terms 
\bea
C_j & = & \sum_{a=1}^3P^{\frac{1}{\sqrt{2}}\left[\ket{(x^1_{a,a})_j} - \ket{(x^2_{a,a})_j}\right]} , 
\label{S32C}
\\
U_{j, j+1}  & = & \sum_{s = 1}^2\sum_{a, b =1 ; a<b}^3~P^{\frac{1}{\sqrt{2}}\left[\ket{\left(x^{s}_{a,b}\right)_j, \left(x^{s}_{b,b}\right)_{j+1}} - \ket{\left(x^{s}_{a,a}\right)_j, \left(x^{s}_{a,b}\right)_{j+1}}\right]},  
\label{S32U}\\
\hspace{-6mm}D_{j, j+1}  & = & \sum_{s = 1}^2\sum_{a, b =1 ; a>b}^3~P^{\frac{1}{\sqrt{2}}\left[\ket{\left(x^{s}_{a,b}\right)_j, \left(x^{s}_{b,b}\right)_{j+1}} - \ket{\left(x^{s}_{a,a}\right)_j, \left(x^{s}_{a,b}\right)_{j+1}}\right]}, 
\label{S32D}\\
  \hspace{-6mm}F^{colored}_{j, j+1} & = &\sum_{s, \, s' = 1}^2 \left[ P^{\sqrt{\frac{2}{3}}\left[\ket{\left(x^{s}_{1,1}\right)_j, \left(x^{s'}_{1,1}\right)_{j+1}} - \frac{1}{2}\left(\ket{\left(x^{s}_{1,2}\right)_j, \left(x^{s'}_{2,1}\right)_{j+1}} + \ket{\left(x^{s}_{1,3}\right)_j, \left(x^{s'}_{3,1}\right)_{j+1}}\right)\right]} \right. \nonumber \\
  &  & \hspace{7mm} \left. +P^{\frac{1}{\sqrt{2}}\left[\ket{\left(x^{s}_{2,2}\right)_j, \left(x^{s'}_{2,2}\right)_{j+1}} - \ket{\left(x^{s}_{2,3}\right)_j, \left(x^{s'}_{3,2}\right)_{j+1}}\right]}\right],
  \label{S32F} \\
\hspace{-6mm}W^{colored}_{j, j+1} & = & \sum_{s, \, s' = 1}^2 \left[P^{\frac{1}{\sqrt{2}}\left[\ket{\left(x^{s}_{1,2}\right)_j, \left(x^{s'}_{2,1}\right)_{j+1}} - \ket{\left(x^{s}_{1,3}\right)_j, \left(x^{s'}_{3,1}\right)_{j+1}}\right]} \right. \nonumber \\ 
&  & \hspace{7mm} \left.+  P^{\frac{1}{\sqrt{2}}\left[\ket{\left(x^{s}_{3,1}\right)_j, \left(x^{s'}_{1,3}\right)_{j+1}} - \ket{\left(x^{s}_{3,2}\right)_j, \left(x^{s'}_{2,3}\right)_{j+1}}\right]}\right],  
\label{S32W}
\eea
and the boundary terms 
\begin{eqnarray}
H_{left}  & = & \sum_{s=1}^2~\left[P^{\ket{\left(x^s_{2,1}\right)_1}} + P^{\ket{\left(x^s_{3,1}\right)_1}} + P^{ \ket{\left(x^s_{3,2}\right)_1}}\right], 
\label{S32left}
\\
H_{right} & = & \sum_{s=1}^2~\left[P^{\ket{\left(x^s_{1,2}\right)_n}}  + P^{\ket{\left(x^s_{1,3}\right)_n}} + P^{\ket{\left(x^s_{2,3}\right)_n}}\right].
\label{S32right}
\end{eqnarray}
(\ref{S32C}) realizes the flat moves (\ref{S32Cmove}). Note that combining (\ref{S32C}) and (\ref{S32U}) (or (\ref{S32D})) 
induces to the independently colored equivalence relations 
\be
x^s_{a,b}\,x^{s'}_{b,b} \sim x^{s'}_{a,a}\, x^s_{a,b} \qquad \mbox{for arbitrary $s$ and $s'$}
\ee
in the case of $a<b$ (or $a>b$). 

The Hamiltonian accompanied with 
\be
H_{bulk,\, disconnected}=\sum_{j=1}^{n-1}\sum_{s,t=1}^2\sum_{a,b,c,d=1; b\neq c}^3~ P^{\ket{\left(x^s_{a,b}\right)_j, \left(x^t_{c,d}\right)_{j+1}}}
\label{S32Hdisc}
\ee
eliminates all the disconnected parts of the paths from the ground states. 

Once again, the Hamiltonian for the connected paths on the reduced Hilbert space can be constructed. 
Details are presented in appendix~\ref{app:Hconnected}. 

\subsubsection{Entanglement entropy for the ground states}

For the first case, the ground states of the Hamiltonian $(\ref{S32H1st}) +(\ref{S32Hdisc})$  
have five-fold degeneracy labelled by the same equivalence classes $\{11\}$, $\{12\}$, $\{21\}$, $\{22\}$ and $\{33\}$ 
as in the $\cS^3_1$ case (with $\lambda=0$). Each equivalence class $\{ab\}$ is described by the length-$n$ SMWs for $\cS^3_1$ with the initial (final) semigroup index $a\, (b)$ 
in which each step is duplicated by the colors $s=1,2$. For example, the length-3 paths from the semigroup index 1 to 1 are 
\bea
& &\hspace{-7mm} P_{3,\,1\to 1} =  (x^1_{1,1} + x^2_{1,1})^3 + \sum_{a=2}^3\Biggl[(x^1_{1,1} + x^2_{1,1})(x^1_{1,a} + x^2_{1,a})(x^1_{a,1} + x^2_{a,1}) \nn \\
& &\hspace{-3mm} + (x^1_{1,a} + x^2_{1,a})(x^1_{a,1} + x^2_{a,1})(x^1_{1,1} + x^2_{1,1}) 
+ (x^1_{1,a} + x^2_{1,a})(x^1_{a,a} + x^2_{a,a})(x^1_{a,1} + x^2_{a,1})\Biggr]. \nonumber \\ 
\eea
As a consequence, the recursion relations satisfied by the SMWs are the same as  (\ref{P11_rec})-(\ref{P21_rec}) with the replacement of $x_{a,b}$ by 
$x^1_{a,b}+x^2_{a,b}$. 
Also, $P_{n,\,3\to 1}=P_{n,\,3\to 1}=0$ and $P_{n,\,3\to 3}=(x^1_{3,3}+x^2_{3,3})^n$. 
The number of the paths of $P_{n,\,a \to b}$ denoted by $N_{n,\, a\to b}$ is obtained by setting all of $x^1_{a,b}$ and $x^2_{a,b}$ to 1 in $P_{n,\,a\to b}$. 
By following the same procedure done in the $\cS^3_1$ case, we see that the generating functions  $N_{a\to b}(x)=\delta_{a,b} + \sum_{n=1}^\infty N_{n,\,a\to b}\,x^n$ 
are given by (\ref{N33g_sol})-(\ref{X}) with $x$ replaced with $2x$ on the r.~h.~s. This implies that the numbers $N_{n,\,a\to b}$ are $2^n$ times the $N_{n,\,a\to b}$ 
in the $\cS^3_1$ case, i.e., 
\be
N_{n,\, a \to b} = 2^n \times (N_{n,\, a\to b} \mbox{ in $\cS^3_1$ case}). 
\ee
It is easy to see that the same relations hold for paths ending at nonzero height: 
\be
N^{(h)}_{n,\, a \to b} = 2^n \times (N^{(h)}_{n,\, a\to b} \mbox{ in $\cS^3_1$ case}). 
\ee
These two lead to the quantities
\be
p^{(h)}_{n,\, a \to b \to c} \equiv \frac{N^{(h)}_{n,\, a\to b}\,N^{(h)}_{n,\,c \to b}}{N_{2n,\,a\to c}}
\ee
being identical to the ones in the $\cS^3_1$ case. Thus, we conclude that the entanglement entropies 
coincide with those obtained in the $\cS^3_1$ case for $\lambda=0$. 
Namely, $S_{A,\, a \to b}$ ($a, b=1,2$) are given by (\ref{EE11f}), and $S_{A,\, 3\to 3}=0$.

\subsection{Hamiltonian for the second case}
\begin{figure}[h!]
\captionsetup{width=0.8\textwidth}
\begin{center}
		\includegraphics[scale=1]{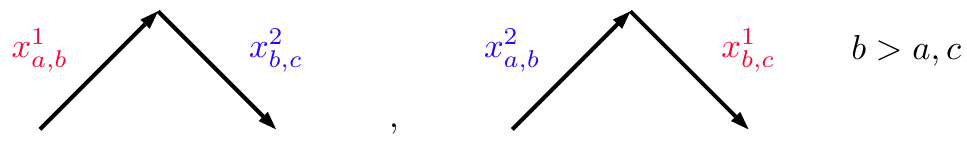} 
	\caption{\small Unmatched up and down steps excluded by (\ref{S32Rbalanced}) from the ground states in the $\cS^3_2$ case. }
\label{leR}
\end{center}
\end{figure}

The Hamiltonian for the second case, called $H^{balanced}$, is given by 
\bea
H^{balanced}& = & \mu \sum_{i=1}^n C_j + \sum_{j=1}^{n-1}~\left[U_{j, j+1} + D_{j, j+1}  + F^{balanced}_{j, j+1} + W^{balanced}_{j, j+1}+R^{balanced}_{j, j+1}\right] \nn \\
& & + H_{left} + H_{right},
\label{S32Hbalanced}
\eea
where $\mu (\geq 0)$ is a tunable parameter introduced for later convenience, 
but we will focus on the case of positive $\mu$ (typically $\mu=1$) unless its value is specified. 
$C_j$, $U_{j,j+1}$, $D_{j,j+1}$, $H_{left}$ and $H_{right}$ are the same as (\ref{S32C})-(\ref{S32D}), (\ref{S32left}) and (\ref{S32right}) respectively, and 
\begin{eqnarray}
\hspace{-6mm} F^{balanced}_{j, j+1} & = &\sum_{s = 1}^2 \left[ P^{\sqrt{\frac23}\left[\ket{\left(x^{s}_{1,1}\right)_j, \left(x^{s}_{1,1}\right)_{j+1}} - \frac{1}{2}\left(\ket{\left(x^{s}_{1,2}\right)_j, \left(x^{s}_{2,1}\right)_{j+1}} + \ket{\left(x^{s}_{1,3}\right)_j, \left(x^{s}_{3,1}\right)_{j+1}}\right)\right]} \right. \nonumber \\
  &  & \hspace{7mm} \left. +P^{\frac{1}{\sqrt{2}}\left[\ket{\left(x^{s}_{2,2}\right)_j, \left(x^{s}_{2,2}\right)_{j+1}} - \ket{\left(x^{s}_{2,3}\right)_j, \left(x^{s}_{3,2}\right)_{j+1}}\right]}\right], 
\label{S32Fbalanced}  
\\
\hspace{-6mm} W^{balanced}_{j, j+1} & = & \sum_{s = 1}^2 P^{\frac{1}{\sqrt{2}}\left[\ket{\left(x^{s}_{1,2}\right)_j, \left(x^{s}_{2,1}\right)_{j+1}} - \ket{\left(x^{s}_{1,3}\right)_j, \left(x^{s}_{3,1}\right)_{j+1}}\right]} \nonumber \\ 
&  &  +\sum_{s,t=1}^2 P^{\frac{1}{\sqrt{2}}\left[\ket{\left(x^{s}_{3,1}\right)_j, \left(x^{t}_{1,3}\right)_{j+1}} - \ket{\left(x^{s}_{3,2}\right)_j, \left(x^{t}_{2,3}\right)_{j+1}}\right]},  
\label{S32Wbalanced}
\\ 
\hspace{-6mm} R^{balanced}_{j, j+1} &= & \sum_{a,b,c=1; \,b> a,c}^3\left[P^{\ket{(x^1_{a,b})_j, (x^2_{b,c})_{j+1}}} + P^{\ket{(x^2_{a,b})_j, (x^1_{b,c})_{j+1}}} \right]. 
\label{S32Rbalanced}
\end{eqnarray}
Notice (\ref{S32Rbalanced}) with (\ref{S32Fbalanced}) and the first line of (\ref{S32Wbalanced}) shows 
that each pair of up and subsequent down steps should be 
matched with respect to the colors in order to gain no energy cost. 
On the other hand, it is not the case for pairs of down and subsequent up steps, as seen from the second line of (\ref{S32Wbalanced}). 
Eq.~(\ref{S32Rbalanced}) excludes unmatched up and down steps shown in Fig. \ref{leR} from the ground states.
%
%

Together with (\ref{S32Hdisc}), the total Hamiltonian 
\be
H^{balanced} + H_{bulk,\,disconnected}
\label{S32Htotal_case2}
\ee
allows the ground states consisting of only the connected paths. 

In appendix~\ref{app:Hconnected}, we present the Hamiltonian for connected paths on the reduced Hilbert space.

\subsubsection{Entanglement entropy for the ground states}
The Hamiltonian for the second case (\ref{S32Htotal_case2}) has five ground states labelled by the same equivalence classes. 
However, the difference from the first case is that the color of up-down steps have to match in each of the walks. 
For example, the length-3 walks from the semigroup index 1 to 1 are 
\bea
& & \hspace{-7mm}P_{3,\,1 \to 1} = (x^1_{1,1} + x^2_{1,1})^3 + \sum_{a=2}^3\Biggl[(x^1_{1,1} + x^2_{1,1})(x^1_{1,a} \,x^1_{a,1} + x^2_{1,a} \,x^2_{a,1}) \nn \\
& & \hspace{-7mm}+ (x^1_{1,a} \,x^1_{a,1} + x^2_{1,a} \,x^2_{a,1}) (x^1_{1,1} + x^2_{1,1}) + x^1_{1,a} (x^1_{a,a} + x^2_{a,a}) x^1_{a,1} +  x^2_{1,a} (x^1_{a,a} + x^2_{a,a}) x^2_{a,1}\Biggr]. 
\nn \\
\eea
Recursion relations for the paths are given by 
\bea
P_{n,\, 1\to 1} & = & \left(\sum_{s=1}^2 x^s_{1,1}\right) P_{n-1,\,1\to 1} +\sum_{s=1}^2 x^s_{1,2}\sum_{i=0}^{n-2}P_{i,\,2 \to 2}\, x^s_{2,1}\,P_{n-2-i,\,1\to 1} \nn \\
& &   +\sum_{s=1}^2 x^s_{1,3}\sum_{i=0}^{n-2}P_{i,\,3 \to 3}\, x^s_{3,1}\,P_{n-2-i,\,1\to 1} 
+\sum_{s=1}^2 x^s_{1,3}\sum_{i=0}^{n-2}P_{i,\,3 \to 3}\, x^s_{3,2}\,P_{n-2-i,\,2\to 1} , \nn \\
\label{S32P11_rec}
\\
P_{n,\, 2\to 2} & = & \left(\sum_{s=1}^2 x^s_{2,2}\right) P_{n-1,\,2\to 2} +\sum_{s=1}^2 x^s_{2,3}\sum_{i=0}^{n-2}P_{i,\,3 \to 3}\, x^s_{3,2}\,P_{n-2-i,\,2\to 2} \nn \\
& & +\sum_{s=1}^2 x^s_{2,3} \sum_{i=0}^{n-2}P_{i,\,3 \to 3}\, x^s_{3,1}\,P_{n-2-i,\,1\to 2}, 
\label{S32P22_rec}
\\
P_{n,\, 2\to 1} & = & \left(\sum_{s=1}^2 x^s_{2,2}\right) P_{n-1,\,2\to 1} +\sum_{s=1}^2 x^s_{2,3}\sum_{i=0}^{n-2}P_{i,\,3 \to 3}\, x^s_{3,2}\,P_{n-2-i,\,2\to 1} \nn \\ 
& & +\sum_{s=1}^2 x^s_{2,3}\sum_{i=0}^{n-2}P_{i,\,3 \to 3}\, x^s_{3,1}\,P_{n-2-i,\,1\to 1}
\label{S32P21_rec}
\eea
with $P_{n,\,3\to 3}=\left(\sum_{s=1}^2 x^s_{3,3}\right)^n$. 
These lead to the recursions among $N_{n,\, a\to b}$ by setting all of the $x_{a,b}^s$ to 1, 
which are solved by introducing the generating functions $N_{a\to b}(x)=\delta_{a,b} + \sum_{n=1}^\infty N_{n,\,a\to b}\,x^n$ as~\footnote{
$N_{n, \, a\to b} = N_{n, \, b\to a}$ holds. Also, $N_{n,\, 3\to a}=N_{n,\, a\to 3}=0$ for $a=1,2$.}  
\bea
N_{1\to 1}(x) & = & \frac{1-4x+2x^2}{4\sqrt{2}\,x^3X}\left[1-\sqrt{1-4X^2}\right], 
\label{S32N11g_sol}
\\
N_{2\to 2}(x) & = & \frac{1-2x}{4x^2(1-4x+2x^2)}\left[1-4x+2x^2- (1-4x-2x^2)\sqrt{1-4X^2}\right], 
\label{S32N22g_sol}
\\
N_{2\to 1}(x) & = & \frac{1}{2\sqrt{2}\,xX}\left[1-\sqrt{1-4X^2}\right]
\label{S32N21g_sol}
\eea
with $N_{n,\,3\to 3}=\frac{1}{1-2x}$ and 
\be 
X\equiv \frac{2\sqrt{2}\,x^3}{(1-2x)(1-4x-2x^2)}. 
\label{S32X}
\ee
(\ref{S32N11g_sol}), (\ref{S32N22g_sol}) and (\ref{S32N21g_sol}) have a singular point 
\be
x_0=\frac{-1+\sqrt{2}}{2}
\label{x0}
\ee
which is the nearest from the origin among their singularities. We can read off the large order behavior of $N_{n, \,a\to b}$ by expanding the generating functions around $x_0$. 
The results are 
\bea
& & N_{n,\,1\to 1}\sim \frac{9}{2^{7/4}x_0^{3/2}\sqrt{\pi}}\,\frac{1}{x_0^n n^{3/2}}, \qquad 
N_{n,\,2\to 2}\sim \frac{1}{2^{7/4}x_0^{3/2}\sqrt{\pi}}\,\frac{1}{x_0^n n^{3/2}},\nn \\
& & N_{n,\,2\to 1}\sim \frac{3}{2^{7/4}x_0^{3/2}\sqrt{\pi}}\,\frac{1}{x_0^n n^{3/2}}
\label{S32N_asym}
\eea 
as $n\to \infty$. 

\paragraph{Paths ending at nonzero height:}

\begin{figure}[h!]
\captionsetup{width=0.8\textwidth}
\begin{center}
		\includegraphics[scale=1]{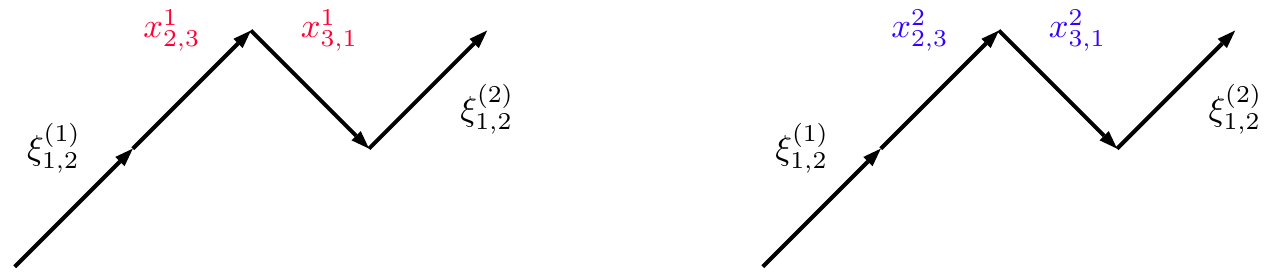} 
	\caption{\small The possible paths in a 4 step walk ending at height 2 in the $\cS^3_2$ case. }
\label{P4}
\end{center}
\end{figure}

For the length-$n$ paths ending at nonzero height $h$ denoted by $P^{(h)}_{n, \,\,a \to b}$, each of the paths has $h$ unmatched 
up-steps. For example, in $P^{(2)}_{4,\, 1\to 2}$ given by
\be
P^{(2)}_{4,\,1\to 2}= (x^1_{1,2} + x^2_{1,2}) (x^1_{2,3} \,x^1_{3,1} + x^2_{2,3} \,x^2_{3,1}) (x^1_{1,2} + x^2_{1,2}) , 
\label{S32_P(2)4}
\ee
the first and last factors, $(x^1_{1,2} + x^2_{1,2})$ are unmatched up-steps. The former goes from height 0 to height 1, and 
the latter from height 1 to 2. The second factor represents matched up and down steps.  
For later convenience, we define by $\tilde{P}^{(h)}_{n,\, a\to b}$ 
the paths in which the color degrees of freedom of unmatched up-steps ($x^1_{c,d}$ and $x^2_{c,d}$) are dropped and 
replaced with a colorless variable $\xi^{(m)}_{c,d}$ ($m=1,\cdots, h$). 
$\xi^{(m)}_{c,d}$ represents an unmatched up-step from the height $m-1$ to $m$. 
In the above example, $\tilde{P}^{(2)}_{4,\,1\to 2}$ becomes 
\be
\tilde{P}^{(2)}_{4,\,1\to 2} = \xi^{(1)}_{1,2} (x^1_{2,3} \,x^1_{3,1} + x^2_{2,3} \,x^2_{3,1}) \xi^{(2)}_{1,2},
\ee
shown in Fig. \ref{P4}.

$\tilde{N}^{(h)}_{n,\,a\to b}$ counts the number of the `paths' (monomials) in $\tilde{P}^{(h)}_{n,\,a \to b}$, which is obtained 
by setting all of the variables 
$x^1_{c,d}$, $x^2_{c,d}$ and $\xi^{(m)}_{c,d}$ to 1 in $\tilde{P}^{(h)}_{n,\, a\to b}$. Then, it is easy to see that the relation  
\be
N^{(h)}_{n, \, a\to b}= 2^h \,\tilde{N}^{(h)}_{n, \, a\to b}
\label{Nh-tildeNh}
\ee
holds. For instance, $N^{(2)}_{4, \,1\to 2}=8$ and $\tilde{N}^{(2)}_{4, \,1\to 2}=2$. 

It is straightforward to write down the recursion relations for $P^{(h)}_{n, \,a \to b}$ and thus for $N^{(h)}_{n,\, a\to b}$. 
The result is given by (\ref{Nh1j_rec}) and (\ref{Nh2j_rec}) 
in which all the terms on the r.~h.~s., except $\delta_{b,3}\delta_{h,1}$, are multiplied by 2, 
and the terms of $\delta_{b,3}\delta_{h,1}$ are multiplied by $2^n$. 
The generating functions defined as (\ref{Nhijg}) are obtained in this case as 
\bea
N^{(h)}_{2\to 1}(x) & = & \frac{1}{2x}\left(\frac{4x^3}{1-4x+2x^2}\,N_{1\to 1}(x)\right)^{h+1}, 
\label{S32N21g}
\\
N^{(h)}_{1\to 1}(x) & = & \frac{1-4x+2x^2}{4x^3}\left(\frac{4x^3}{1-4x+2x^2}\,N_{1\to 1}(x)\right)^{h+1}, 
\label{S32Nh11g} 
\\
N^{(h)}_{2\to 2}(x) & = & \left(\frac{4x^3}{1-4x+2x^2}\,N_{1\to 1}(x)\right)^h N_{2\to 2}(x), 
\label{S32N22g}
\\
N^{(h)}_{1\to 2}(x) & = & \frac{1-4x+2x^2}{2x^2}\left(\frac{4x^3}{1-4x+2x^2}\,N_{1\to 1}(x)\right)^h N_{2\to 2}(x), 
\label{S32N12g}
\\
N^{(h)}_{2\to 3}(x) & = & \frac{1}{2x^2}\left(\frac{4x^3}{1-4x+2x^2}\,N_{1\to 1}(x)\right)^h \left[1-2x-2x^2N_{2\to 2}(x)\right], 
\label{S32N23g}
\\
N^{(h)}_{1\to 3}(x) & = & -\delta_{h,1}\frac{1}{x}+\frac{1-4x+2x^2}{4x^4}\left(\frac{4x^3}{1-4x+2x^2}\,N_{1\to 1}(x)\right)^h
\nn \\
& & \times \left[1-2x-2x^2N_{2\to 2}(x)\right]
\label{S32N13g}
\eea
for $h\geq 1$. 
Similarly to the $\cS^3_1$ case, the use of the identity (\ref{Id}) with (\ref{S32X}) provides the asymptotic behavior of the coefficients 
$N^{(h)}_{n, \, a\to b}$ relevant to the computation of the entanglement entropies. 
Here, we have 
\be
N^{(h)}_{p,\,2\to 1}=2^{\frac12h-1}\sum_{n,\ell\geq 0}^* \sum_{j=0}^\ell N^{(h)}_{n}
\binomi{p-2n-\ell-j-2}{n} \binomi{n+\ell}{\ell} \binomi{\ell}{j}2^{p-\frac32n+\ell-2j},
\ee
where the asterisk (${}^*$) put to the sum means again that $n$ and $\ell$ run satisfying $p-3n-\ell-j\geq 2$.  
The saddle point of the summand is found to be 
\be
n\sim \frac{2\sqrt{2}}{9}x_0 p, \qquad \ell\sim \frac19(2+3\sqrt{2})p, \qquad j\sim \frac{2\sqrt{2}}{9}x_0 p 
\ee 
for $p$ large. Repeating what was manipulated in the $\cS^3_1$ case, we end up with
\bea
N^{(h)}_{p,\,2\to 1} & \sim & 2^{\frac12h}(h+1)\,e^{-\frac{9}{4\sqrt{2}x_0p}(h+1)^2}\,\frac{3}{2^{7/4}x_0^{3/2}\sqrt{\pi}}\,
\frac{1}{x_0^p\,p^{3/2}} , 
\label{S32Nh21_asym}
\\
N^{(h)}_{p,\,1\to 1} & \sim & 2^{\frac12h}(h+1)\,e^{-\frac{9}{4\sqrt{2}x_0p}(h+1)^2}\,\frac{9}{2^{7/4}x_0^{3/2}\sqrt{\pi}}\,
\frac{1}{x_0^p\,p^{3/2}} , 
\label{S32Nh11_asym}
\\
N^{(h)}_{p,\,1\to 2} & \sim & 2^{\frac12h}\left[ 2h\,e^{-\frac{9}{4\sqrt{2}x_0p}h^2} + (h+1)\,e^{-\frac{9}{4\sqrt{2}x_0p}(h+1)^2}\right]
\,\frac{3}{2^{7/4}x_0^{3/2}\sqrt{\pi}}\,\frac{1}{x_0^p\,p^{3/2}} , 
\label{S32Nh12_asym}
\\
N^{(h)}_{p,\,2\to 2} & \sim & 2^{\frac12h}\left[ 2h\,e^{-\frac{9}{4\sqrt{2}x_0p}h^2} + (h+1)\,e^{-\frac{9}{4\sqrt{2}x_0p}(h+1)^2}\right]
\,\frac{1}{2^{7/4}x_0^{3/2}\sqrt{\pi}}\,\frac{1}{x_0^p\,p^{3/2}} , 
\label{S32Nh22_asym}
\\
N^{(h)}_{p,\,2\to 3} & \sim & 2^{\frac12h}\left[ 4h\,e^{-\frac{9}{4\sqrt{2}x_0p}h^2} - (h+1)\,e^{-\frac{9}{4\sqrt{2}x_0p}(h+1)^2}\right]
\,\frac{1}{2^{7/4}x_0^{3/2}\sqrt{\pi}}\,\frac{1}{x_0^p\,p^{3/2}} , 
\label{S32Nh23_asym}
\\
N^{(h)}_{p,\,1\to 3} & \sim & 2^{\frac12h}\left[ 4h\,e^{-\frac{9}{4\sqrt{2}x_0p}h^2} - (h+1)\,e^{-\frac{9}{4\sqrt{2}x_0p}(h+1)^2}\right]
\,\frac{3}{2^{7/4}x_0^{3/2}\sqrt{\pi}}\,\frac{1}{x_0^p\,p^{3/2}}, 
\label{S32Nh13_asym}
\eea
up to multiplicative factors of $\left[1+ O\left(p^{-1}\right)\right]$. 

\paragraph{Schmidt decomposition and entanglement entropy:}

Let us consider dividing the length-$2n$ paths $P_{2n,\, a\to c}$ into two at the midpoint. 
$P^{(0\to h)}_{n,\, a\to b}\equiv P^{(h)}_{n,\, a \to b}$ is the sum of the length-$n$ paths starting at height 0 and ending at height $h$. 
$P^{(h\to 0)}_{n,\, b \to a}$ represents the reversed paths starting at height $h$ and ending at height 0. 
Then, $P^{(0\to h)}_{n, \, a\to b}$ has $h$ unmatched up-steps, 
while $P^{(h\to 0)}_{n,\, b\to a'}$ has the same number of unmatched down-steps. 
As $\tilde{P}^{(0\to h)}_{n,\,a\to b}\equiv \tilde{P}^{(h)}_{n,\, a \to b}$ defined before, 
we define by $\tilde{P}^{(h\to 0)}_{n,\, b\to a'}$ the paths in which 
the color degrees of freedom of $h$ unmatched down-steps ($x^1_{c,d}$ and $x^2_{c,d}$) are dropped 
and replaced by colorless variables $\xi^{(m)}_{c,d}$ ($m=1,\cdots, h$). 
Namely, $\tilde{P}^{(h\to 0)}_{n,\, b\to a}$ is the reversed paths of $\tilde{P}^{(0\to h)}_{n,\,a\to b}$. 
In considering the division of $P_{2n,\,a\to c}$ into the two parts, unmatched up-steps in the left half paths and 
unmatched down-steps in the right half paths are matched with each other. Thus, we can express the division as 
\be
P_{2n,\, a\to c}=\sum_{h\geq 0}\sum_{b=1}^3\sum_{\xi^{(1)}=x^1,\, x^2}\cdots \sum_{\xi^{(h)}=x^1,\, x^2} \tilde{P}^{(0\to h)}_{n,\, a \to b} \,
\tilde{P}^{(h\to 0)}_{n, \, b\to c}. 
\label{S32_division}
\ee 
For the numbers of the paths $\tilde{P}^{(0\to h)}_{n,\,a\to b}$ and $\tilde{P}^{(h\to 0)}_{n,\, b\to c}$ denoted by 
$\tilde{N}^{(0\to h)}_{n,\,a\to b}$ and $\tilde{N}^{(h\to 0)}_{n,\, b\to c}$ respectively, 
\be
\tilde{N}^{(h\to 0)}_{n,\, b\to c}=\tilde{N}^{(0\to h)}_{n,\, c\to b}=2^{-h}N^{(h)}_{n,\, c\to b}
\label{tildeNh-Nh}
\ee
holds from (\ref{Nh-tildeNh}). Then, (\ref{S32_division}) with (\ref{tildeNh-Nh}) leads to the following composition law: 
\bea
N_{2n\, a\to c} & = & \sum_{h=0}^\infty 2^h \sum_{b=1}^3\tilde{N}^{(0\to h)}_{n,\, a\to b}\,\tilde{N}^{(h\to 0)}_{n,\, b \to c} \nn \\
 & = & \sum_{h=0}^\infty 2^{-h} \sum_{b=1}^3 N^{(h)}_{n,\, a\to b}\,N^{(h)}_{n,\, c \to b}. 
 \label{S32composition}
\eea  
We can check (\ref{S32Nh21_asym})-(\ref{S32Nh13_asym}) satisfying (\ref{S32composition}) up to $O(1/p)$ errors. 

The normalized ground state in the sector $\{11\}$ for the system of length $2n$ is expressed as in (\ref{GS11}). 
However, due to the match of the steps across the midpoint (\ref{S32_division}), 
the Schmidt decomposition by the states of the length-$n$ subsystems A and B is expressed as a different form 
from (\ref{GShi1}) and (\ref{GS11decomp}): 
\be
\ket{P_{2n,\,1\to 1}} = \sum_{h\geq 0}\sum_{a=1}^3\sum_{\xi^{(1)}=x^1,\,x^2}\cdots \sum_{\xi^{(h)}=x^1,\, x^2} 
\sqrt{p^{(h)}_{n,\,1\to a \to 1}}\,\ket{\tilde{P}^{(0\to h)}_{n,\, 1 \to a}}\otimes
\ket{\tilde{P}^{(h\to 0)}_{n, \, a\to 1}}
\label{S32_Schmidt}
\ee 
with 
\be
\ket{\tilde{P}^{(0\to h)}_{n,\,1\to a}}\equiv \frac{1}{\sqrt{\tilde{N}^{(h)}_{n,\,1\to a}}}\sum_{w\in\tilde{P}^{(0\to h)}_{n,\, 1\to a}}\ket{w}, 
\qquad 
\ket{\tilde{P}^{(h\to 0)}_{n,\,1\to a}}\equiv \frac{1}{\sqrt{\tilde{N}^{(h)}_{n,\,1\to a}}}\sum_{w\in\tilde{P}^{(h\to 0)}_{n,\, 1\to a}}\ket{w}. 
\ee
The coefficient is given by 
\be
p^{(h)}_{1\to a \to 1}\equiv\frac{\left(\tilde{N}^{(h)}_{n, 1\to a}\right)^2}{N_{2n,\, 1\to 1}}
=2^{-2h}\frac{\left(N^{(h)}_{n, 1\to a}\right)^2}{N_{2n,\, 1\to 1}}. 
\label{S32pcoeff}
\ee
Then, the reduced density matrix for the subsystem A takes the diagonal form as  
\bea
\rho_{A,\, 1\to 1} & = & \Tr_B\ket{P_{2n,\,1\to 1}}\bra{P_{2n,\,1\to 1}} \nn \\
& = & \sum_{h\geq 0}\sum_{a=1}^3\sum_{\xi^{(1)}=x^1,\,x^2}\cdots \sum_{\xi^{(h)}=x^1,\, x^2} p^{(h)}_{1\to a \to 1}
\ket{\tilde{P}^{(0\to h)}_{n,\,1\to a}}\bra{\tilde{P}^{(0\to h)}_{n,\,1\to a}}, 
\label{S32rhoA}
\eea
from which the entanglement entropy reads 
\bea
S_{A,\,1\to 1} & = & -\sum_{h\geq 0}\sum_{a=1}^3\sum_{\xi^{(1)}=x^1,\,x^2}\cdots \sum_{\xi^{(h)}=x^1,\, x^2} 
p^{(h)}_{1\to a \to 1}\ln p^{(h)}_{1\to a \to 1}
\nn \\
& = & -\sum_{h\geq 0}\sum_{a=1}^3 2^h\,
p^{(h)}_{1\to a \to 1}\ln p^{(h)}_{1\to a \to 1}. 
\label{S32EE11}
\eea
Here, since $p^{(h)}_{1\to a \to 1}$ is independent of $\xi^{(m)}$ ($m=1,\cdots, h$), the sum over them yields the factor $2^h$. 

By using (\ref{S32N_asym}), (\ref{S32Nh11_asym}), (\ref{S32Nh12_asym}) and (\ref{S32Nh13_asym}), straightforward computation 
leads to
\bea
S_{A,\,1\to 1} & = & (2\ln 2)\sqrt{\frac{2\sigma n}{\pi}}+ \frac12\ln n +\frac12\ln(2\pi \sigma)+\gamma-\frac12
+\ln \frac{3}{2^{1/3}} \nn \\
& & +(\mbox{terms vanishing as $n\to \infty$})
\label{S32EE11f}
\eea
with $\sigma\equiv \frac{\sqrt{2}\,x_0}{9}=\frac{-1+\sqrt{2}}{9\sqrt{2}}$. 
The first term provides the violation of the area law by the square root of the volume 
that has much stronger entanglement than the first case of the logarithmic violation. 
Similar to the colored Motzkin walks~\cite{shor}, 
(\ref{S32pcoeff}) has a factor $2^{-h}$ that is crucial to the behavior of the square root of the volume.  
The expression (\ref{S32EE11f}) is similar to the colored ($s=2$) Motzkin walks in~\cite{shor}, 
but there are some differences. The value of $\sigma$ is not the same 
as $\frac{\sqrt{2}}{2\sqrt{2}+1}$ for the $s=2$ Motzkin walks, 
and the last term in the constant contributions ($\ln \frac{3}{2^{1/3}}$) 
is additional in our case, and would reflect the structure of the semigroup $\cS^3_2$. 
For the other ground states in the sectors $\{12\}$, $\{21\}$, $\{22\}$, the same result is obtained, while 
the entanglement entropy vanishes for the sector of $\{33\}$.  
 
\paragraph{Quantum phase transition:}
For arbitrary positive $\mu$, 
the structure of the ground states never changes and the result of the entanglement entropies is the same as 
in the above. 

However, at $\mu=0$ (i.e., in the absence of the $C_j$ terms), the equivalence relations changing the colors of the steps disappear, i.e. a given coloring pattern through a path never changes under the equivalence moves. 
At the boundaries of islands of homogeneous color in the path, the heights are fixed, but semigroup indices can change between 1 and 2. 
This is due to the the second term of $W^{balanced}$ in (\ref{S32Wbalanced}).  
For example, consider a flat path of length 10; 
$
(x^1_{1,1})^3\,(x^2_{1,1})^2\,(x^1_{1,1})^5. 
$
The equivalence moves amount to the paths 
\be
\sum_{a,b=1}^2\left(P_{3,\, 1\to a}\left|_{x^1}\right.\right) \left(P_{2,\, a\to b}\left|_{x^2}\right.\right) \left(P_{5,\, b\to 1}\left|_{x^1}\right.\right), 
\ee
where $P_{n, a \to b}\left|_{x^1}\right.$ ($P_{n, a \to b}\left|_{x^2}\right.$) denotes the length-$n$ $\cS^3_1$ SMWs 
with the single color $x^1$ ($x^2$).  
For the length-$n$ system, 
the original ground state in the  $\{ab\}$ ($a,b=1,2$) equivalence class gets split into lots of equivalence classes that are labelled by 
$(h_{i_1},\,h_{i_2}, \cdots, h_{i_k})$ and the color of the first step. 
$(i_1, h_{i_1}), (i_2,h_{i_2}), \cdots, (i_k,h_{i_k})$ represent the boundary points of the islands in the path in an $(x,y)$-plane, where 
$k=0, 1, \cdots, n-1$, and $0<i_1<i_2<\cdots<i_k$. 
The number of the equivalence classes--i.e. the number of ground states--grows exponentially as $c^n$ with 
$
2< c< \frac{1}{x_0}=\frac{2}{-1+\sqrt{2}}=4.828\ldots .    
$
The lower bound is from the restricted cases $h_{i_1}=h_{i_2}=\cdots =h_{i_k}=0$, and the upper bound is set by the total number of the 
paths (\ref{S32N_asym}). 
In such highly degenerate ground states, states with a totally homogeneous color (totally $x^{(1)}$ or totally $x^{(2)}$) will give the maximal entanglement entropy, 
which reduces to the case of $\cS^3_1$ with $\lambda=0$ behaving as $\ln n$. 

We thus find that there is a quantum phase transition between the entanglement entropy behaving as $\ln n$ at $\lambda=0$ and 
that behaving as $\sqrt{n}$ at any positive $\mu$, depicted in Fig.~\ref{phase1}.

\section{The model based on $\cS^2_1$} \label{sec:s21}

It is easy to see that we cannot go beyond height 1 in case of the $\cS^2_1$ semigroup, and 
as a result we do not expect many correlations between the two halves of the chain, giving us an area law for the entanglement entropy in the subsystem size.

\begin{figure}[h]
\captionsetup{width=1\textwidth}
\centering
\includegraphics[scale = 0.8]{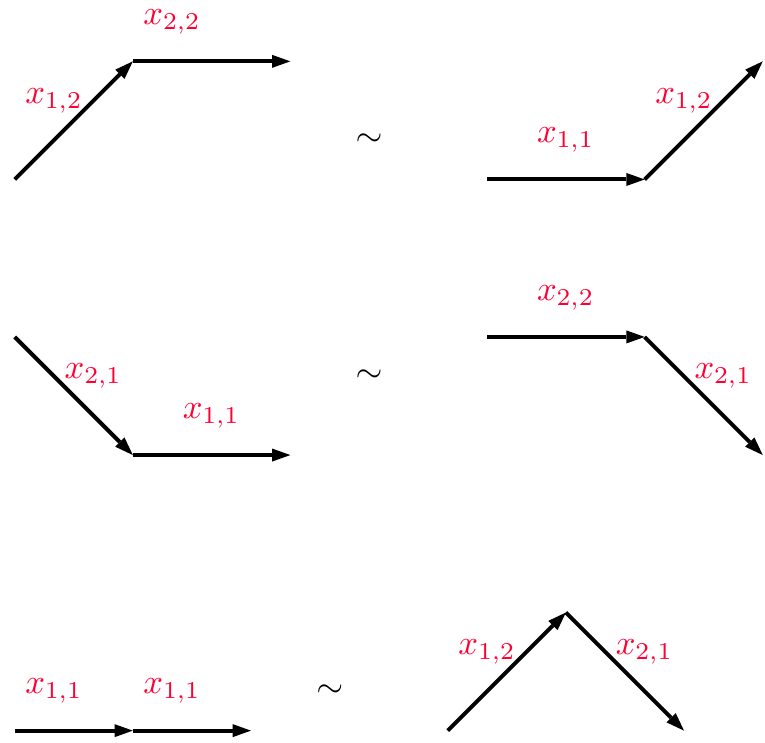}
\caption{\small The local equivalence moves for the case of the SIS $\cS^2_1$. } 
\label{les21}
\end{figure}
The local equivalence moves are shown in Fig. \ref{les21}. 
These give rise to the frustration-free Hamiltonian 
\begin{equation}
H = H_{left} + H_{bulk} + H_{right},
\label{S21H}
\end{equation}
with
\begin{equation}
H_{bulk} = \sum_{j=1}^{n-1} \left(U_{j, j+1} + D_{j, j+1} + F_{j, j+1} \right),
\end{equation}
where
\begin{eqnarray}
U_{j, j+1} & = & P^{\frac{1}{\sqrt{2}}\left[\ket{\left(x_{1,2}\right)_j, \left(x_{2,2}\right)_{j+1}} - \ket{\left(x_{1,1}\right)_j, \left(x_{1,2}\right)_{j+1}}\right]}, \\  
D_{j, j+1} & = & P^{\frac{1}{\sqrt{2}}\left[\ket{\left(x_{2,1}\right)_j, \left(x_{1,1}\right)_{j+1}} - \ket{\left(x_{2,2}\right)_j, \left(x_{2,1}\right)_{j+1}}\right]}, \\
F_{j, j+1} & = & P^{\frac{1}{\sqrt{2}}\left[\ket{\left(x_{1,1}\right)_j, \left(x_{1,1}\right)_{j+1}} - \ket{\left(x_{1,2}\right)_j, \left(x_{2,1}\right)_{j+1}}\right]}, 
\end{eqnarray}
and 
\begin{eqnarray}
H_{left} & = & P^{\ket{\left(x_{2,1}\right)_1}}, \\ 
H_{right} & = & P^{\ket{\left(x_{1,2}\right)_n}}.
\end{eqnarray}
As in the $\cS^3_1$ case this Hamiltonian contains an extensive GSD due to the presence of disconnected and partially connected paths. 
These are lifted out of the ground states by adding 
\be
H_{bulk, \,disconnected} = \sum_{j=1}^{n-1}\sum_{a,b,c, d = 1; b\neq c}^2~ P^{\ket{\left(x_{a,b}\right)_j, \left(x_{c,d}\right)_{j+1}}}
\ee
to the Hamiltonian (\ref{S21H}). Then, the ground states correspond to the connected 
length-$n$ $\cS^2_1$ SMWs. 

We can also focus on the connected paths by again going to the reduced Hilbert space 
and writing down a Hamiltonian for the connected part, $H_{connected}$. This is given by
\begin{equation}
H_{connected} = H_{left, connected} + H_{bulk, connected} + H_{right, connected},
\end{equation}
where
\begin{equation}
H_{bulk, connected} = \sum_{i=1}^{n-1} \left({\cal U}_i + {\cal D}_i + {\cal F}_i \right),
\end{equation}
with $i$ denoting the site index and 
\begin{eqnarray}
{\cal U}_i & = & P^{\ket{1}}_{i-1}\,\frac12\left(1^{12}_i - X^{12}_i\right) P^{\ket{2}}_{i+1}, \\
{\cal D}_i & = & P^{\ket{2}}_{i-1}\,\frac12\left(1^{12}_i - X^{12}_i\right) P^{\ket{1}}_{i+1}, \\
{\cal F}_i & = & P^{\ket{1}}_{i-1}\,\frac12\left(1^{12}_i - X^{12}_i\right) P^{\ket{1}}_{i+1},
\end{eqnarray}
and 
\begin{eqnarray}
H_{left, connected} & = & P^{\ket{2}}_0 P^{\ket{1}}_1, \\
H_{right, connected} & = & P^{\ket{1}}_{n-1} P^{\ket{2}}_n.
\end{eqnarray}

The ground state space of this Hamiltonian is far simpler than the $\cS^3_1$ case, as we only form two equivalence classes that cannot be mapped into each other, namely $\{11\}$ and $\{22\}$, with the latter being a simple product state with $x_{2,2}$ on every step. It is easy to check that this is indeed the case as we no longer have the equivalence class $\{12\}$ as it is impossible to come down and end in the semigroup index 2 at the end of the chain. 

The normalization of the $\{11\}$ case is easy to compute by solving the recursion relation
\begin{equation}
N_{1\to 1}(x) - 1 = xN_{1\to 1}(x) + x^2N_{2\to 2}(x)N_{1\to 1}(x),
\end{equation}
by noting that $N_{2\to 2}(x) = \frac{1}{1-x}$ resulting in 
\begin{equation}
N_{1\to 1}(x) = \frac{1-x}{1-2x}.
\end{equation}
The behavior of the coefficients of this term reads 
\begin{equation}
N_{n,\,1\to 1} = 2^{n-1}  \qquad (\textrm{for}~n\geq 1).
\end{equation}

To compute the entanglement entropy we also need to know the number of paths to height $h$ when we perform the Schmidt decomposition. 
In this case it is easy to see that we cannot go beyond height 1 due to the algebra of $\cS^2_1$ and that the only set of paths that can do this is captured by $P^{(1)}_{n, \, 1\to 2}$. The number of such paths is computed by using the recursion relation
\begin{equation}
N^{(1)}_{1\to 2}(x) = x N^{(1)}_{1\to 2}(x) + xN^{(0)}_{2\to 2}(x) + x^2 N^{(0)}_{2\to 2}(x)N^{(1)}_{1\to 2}(x),
\end{equation}
and $N^{(0)}_{2\to 2}(x) = \frac{1}{1-x}$ to give
\begin{equation}
N^{(1)}_{1\to 2}(x) = \frac{x}{1-2x},
\end{equation}
from which its coefficient behaves as 
\begin{equation}
N^{(1)}_{n,\,1\to 2} = 2^{n-1}  \qquad (\textrm{for}~n\geq 1).
\end{equation}

We can now compute the entanglement entropy of the ground state $\{11\}$ as
\begin{equation}
S_{A, 1\to1} = -\left[ p^{(0)}_{n,\, 1\to 1 \to 1}\ln p^{(0)}_{n,\, 1\to 1 \to 1} + p^{(1)}_{n,\, 1\to 2 \to 1}\ln p^{(1)}_{n,\, 1\to 2 \to 1}\right],
\end{equation}
with $p^{(0)}_{n,\, 1\to 1 \to 1} = p^{(1)}_{n,\, 1\to 2 \to 1} = \frac{1}{2}$ giving a constant entanglement entropy of $S_{A, 1\to1} = \ln 2$. This is expected, as this system has weak correlations between the two halves of the chain. 

\section{Discussion and Outlook}\label{outlook} 
\setcounter{equation}{0}

So far we have considered the SMWs on open chains and seen that there are five ground states in the 
$\cS^3_1$ ($\lambda=0$) and $\cS^3_2$ ($\mu> 0$) 
corresponding to the equivalence classes in the semigroup indices. The situation changes when we close the chain, 
as we lose two of the five equivalence classes, $\{12\}$ and $\{21\}$, as the endpoints of the open chain have to be identified. 
(This holds when the same Hamiltonians as in the open case are used.) 
We are then left with three ground states on the closed chain which cannot be mapped into each other by local unitary operators. 
Similarly to what we discussed in Sec.~\ref{l0} in the open chains, this GSD is sensitive to local perturbations at the site 
$i=0(\equiv n)$, while it is stable for local perturbations at other sites. For the case in which the Hamiltonian only has the bulk terms acting on all the links/sites
of the closed chain (that is $H_{left}$ and $H_{right}$ from the open chain Hamiltonian are not included), we can see that there are two equivalence classes, $\{11\}$ and $\{22\}$ from the previous case, 
which combine into a single equivalence class
and $\{33\}$ forms a separate equivalence class. This GSD of 2 is now stable against local perturbations that commute with the local equivalence moves.


%

There are several directions one can take from this work. An interesting possibility would be to study the excited states and the scaling of the spectral gap in these systems and check for many-body localization-like states in the high-energy sector of these models. 
We can construct the systems out of higher SISs, $\cS^n_p$ naturally from this framework. With the introduction of higher semigroups we can reach bigger heights and we expect to obtain more correlations between the two halves of the chain with this. Finally we can study the continuum versions of this model as done in  \cite{new} and check whether there are emerging symmetries in this system.

\section*{Acknowledgements}
This work is supported by IBS-R018-D2. PP thanks T. R. Govindarajan, Israel Klich and Hal Tasaki for useful discussions. 

\appendix
\section{Derivation of (\ref{Id})}
\label{app:Dyck}
\setcounter{equation}{0}
In this appendix, we consider the DWs on a two-dimensional plane $(x,y)$ in two different ways and derive the identity (\ref{Id}). 
The DWs are composed of an up and a down move. The walks start at the origin $(0,0)$, and never enter the region $y<0$. 
For the case where the walks end at $(2n,0)$, the number of such walks $N_{2n}$ satisfies the recursion: 
\be
N_{2n}=\sum_{i=0}^{n-1}N_{2i}N_{2n-2i-2} \qquad (n\geq 1), 
\ee 
which can be solved by using the generating function $N(X)=\sum_{m=0}^\infty N_{2m}X^{2m}$ with $N_0=1$ as 
\be
N(X)=\frac{1}{2X^2}\left(1-\sqrt{1-4X^2}\right). 
\label{Ng}
\ee 
For the case that the walks end at a positive height $(n,h)$, the recursion relations for the number of the walks $N^{(h)}_{n}$ are 
\be
N^{(2k)}_{2m}=N^{(2k-1)}_{2m-1}+\sum_{i=0}^{m-1}N_{2i}N^{(2k)}_{2m-2i-2} 
\label{Neven_rec}
\ee
for $n$ and $h$ even ($n=2m$, $h=2k$), and 
\be
N^{(2k-1)}_{2m+1}=N^{(2k-2)}_{2m}+\sum_{i=0}^{m-1}N_{2i}N^{(2k-1)}_{2m-2i-1}
\label{Nodd_rec}
\ee
for $n$ and $h$ odd ($n=2m+1$, $h=2k-1$). Note that $N^{(h)}_n$ vanishes when $n+h$ odd. By introducing the generating function
\be
N^{(h)}(X)=\sum_{n=0}^\infty N^{(h)}_nX^n
\ee
with $N^{(h)}_0=0$ for $h\geq1$, the recursions (\ref{Neven_rec}) and (\ref{Nodd_rec}) can be solved as 
\be
N^{(h)}(X)=X^hN(X)^{h+1}. 
\label{Nh_sol}
\ee

On the other hand, as presented in the literature (for instance, \cite{dyck, shor}), $N^{(h)}_n$ is given as a solution of the Ballot problem by
\be
N^{(h)}_n =\frac{h+1}{n+1}\binomi{n+1}{\frac{n-h}{2}}=\frac{h+1}{\frac{n+h}{2}+1} \binomi{n}{\frac{n+h}{2}}. 
\label{Nh_sol2}
\ee

Combining (\ref{Ng}), (\ref{Nh_sol}) and (\ref{Nh_sol2}) leads to the identity (\ref{Id}).  

\section{Fluctuations around the saddle point}
\label{app:fluctuation}
\setcounter{equation}{0}
In this appendix, we show that fluctuations of $e^{-\frac{1}{2n}(h+1)^2}$ around the saddle point (\ref{saddle}) can be neglected 
in the sum (\ref{Nh21_sol}):
\be
N^{(h)}_{p,\,2 \to 1}\simeq (h+1)\sum_{n,\ell\geq 0}^*\sum_{j=0}^\ell
e^{-\frac{1}{2n}(h+1)^2}\,N_n^{(0)}\binomi{p-2n-\ell-j-2}{n}\binomi{n+\ell}{\ell}\binomi{\ell}{j}2^\ell. 
\label{Nh21app1}
\ee
Let us consider the case of $h$ even ($h=2k$), meaning that $n$ should also run over even integers ($n=2m$). We put 
\be
m=\frac{1}{27}p+x, \qquad \ell=\frac{16}{27}p+y, \qquad j=\frac{4}{27}p+z,
\ee
where $x$, $y$ and $z$ stand for fluctuations around the saddle point and $x, y, z\ll p$. 
Then, the exponential factor is expanded as 
\be
e^{-\frac{1}{2n}(h+1)^2}= e^{-\frac{27}{4p}(2k+1)^2}\times \exp\left[+\frac{27^2}{4p^2}(2k+1)^2x -\frac{27^3}{4p^3}(2k+1)^2x^2+\cdots\right].
\label{expfactor}
\ee

After a straightforward calculation, we see that (\ref{Nh21app1}) can be expressed in the form 
of integrals around the saddle point:
\bea
N_{p,\,2\to 1}^{(2k)} & \simeq & (2k+1) e^{-\frac{27}{4p}(2k+1)^2}\frac{3^8}{2^4\pi^2}\frac{3^p}{p^3}
\int_{-\infty}^{\infty}dx\,dy\,dz\,e^{-\frac{1}{p}f_2(x,y,z)} \nn \\
& & \times \exp\left[-\frac{1}{p}\left(255x+\frac{219}{8}y+\frac{117}{4}z+18\right) +O\left(\frac{x^2}{p^2},\,\frac{x^3}{p^2}\right)\right]
\nn \\
& & \times \exp\left[+\frac{27^2}{4p^2}(2k+1)^2x -\frac{27^3}{4p^3}(2k+1)^2x^2+\cdots\right],
\label{Nh21app2}
\eea
where $f_2(x,y,z)$ is a positive-definite quadratic form of $x$, $y$ and $z$ given by 
\be
f_2(x,y,z)\equiv 465x^2+\frac{75}{8}y^2+\frac{27}{2}z^2+123xy+\frac{63}{4}yz+126xz. 
\ee
The integrals in (\ref{Nh21app2}) can be computed as Gaussian integrals 
by bringing down the factors in the second and third lines from the exponentials. 
Note that $x$, $y$ and $z$ are effectively at most $O(\sqrt{p})$, and contributions from the linear terms in $x$, $y$ and $z$ vanish 
due to the parity.   
Then, it is easy to see that the fluctuations provide the contribution of the order at most $O(p^{-1})$, and can be safely neglected. 

For the case of $h$ odd, we can argue in parallel and obtain the same conclusion.

\section{$H_{connected}$ in the case of $\cS^3_2$}
\label{app:Hconnected}
\setcounter{equation}{0}
In this appendix, we discuss the reduced Hilbert space that just describes the connected paths, 
and construct the Hamiltonians in terms of degrees of freedom at sites for the two cases in the semigroup $\cS^3_2$. 

The original Hilbert space on the links has a total dimension $18^n$, with $n$ the size of the one-dimensional chain. 
The total dimension of the reduced Hilbert space is $6^{n+1}$. 
This is easily seen to be the case as the local Hilbert space on each site has three choices for the semigroup index, $\{\ket{1}, \ket{2}, \ket{3}\}$ 
and two choices for the color degree of freedom, $s\in \{1, 2\}$. 
We can now write down the basis of this six-dimensional Hilbert space as $\{\ket{1^1}, \ket{2^1}, \ket{3^1}, \ket{1^2}, \ket{2^2}, \ket{3^2}\}$. 

As a new ingredient that does not appear in the cases of $\cS^3_1$ and $\cS^2_1$, 
while the color degrees of freedom are put on the links in the original Hilbert space, they are assigned at the sites in the 
reduced Hilbert space. 
Let us assign the colors on the link $(j, j+1)$ in the original Hilbert space to the site $j+1$ in the reduced one. 
Then, the color degrees of freedom at the site 0 become surplus. 
For example, a state of the length-4 chain $\ket{(x^{(2)}_{1,2})_1, \,(x^{(2)}_{2,3})_2,\,(x^{(1)}_{3,1})_3}$ will be expressed as 
$\ket{1^s_0, \,2^2_1, \,3^2_2,\, 1^1_3}$ with the color $s$ undetermined. 
Here, we take into account both of the possibilities $s=1$ and $s=2$, making an additional 2-fold degeneracy for each connected path 
of the random walks in the original description. 
After obtaining all of the length-$n$ paths eventually, the doubling is resolved by simply identfying the states 
$\ket{a^1_0, \,\cdots}$ and $\ket{a^2_0, \,\cdots}$. 

\subsection{$H_{connected}$ in the first case}
The bulk terms (\ref{S32C})-(\ref{S32F}) are expressed by the site variables on the reduced Hilbert space as 
\bea
\hspace{-7mm} {\cal C}_i & = & \sum_{s=1}^2\sum_{a=1}^3 P_{i-1}^{\ket{a^s}}\,\frac12\left(1^{a^1a^2}_i - X^{a^1a^2}_i\right), 
\\
\hspace{-7mm} {\cal U}_i & = & \sum_{s, t =1}^2\sum_{a,b=1 ; a<b}^3~P^{\ket{a^{s}}}_{i-1}\,\frac12\left(1^{a^{t}b^{t}}_i-X^{a^{t}b^{t}}_i\right)
P^{\ket{b^{t}}}_{i+1},  
\\
\hspace{-7mm} {\cal D}_i & = & \sum_{s, t =1}^2\sum_{a,b=1 ; a>b}^3~P^{\ket{a^{s}}}_{i-1}\,\frac12\left(1^{a^{t}b^{t}}_i-X^{a^{t}b^{t}}_i\right)
P^{\ket{b^{t}}}_{i+1},
\\ 
\hspace{-7mm} {\cal F}^{colored}_i & = & \sum_{s, t, u =1}^2 \left[P^{\ket{1^{s}}}_{i-1}\,\frac{1}{3}\left(1^{1^{t}2^{t}}_i-X^{1^{t}2^{t}}_i\right)P^{\ket{1^{u}}}_{i+1} 
+  P^{\ket{1^{s}}}_{i-1}\,\frac{1}{3}\left(1^{1^{t}3^{t}}_i-X^{1^{t}3^{t}}_i\right)P^{\ket{1^{u}}}_{i+1} \right. \nonumber \\
              & &\hspace{7mm}\left.  - P^{\ket{1^{s}}}_{i-1}\,\frac{1}{6}\left(1^{2^{t}3^{t}}_i-X^{2^{t}3^{t}}_i\right)P^{\ket{2^{u}}}_{i+1}  
               +  P^{\ket{2^{s}}}_{i-1}\,\frac12\left(1^{2^{t}3^{t}}_i-X^{2^{t}3^{t}}_i\right)P^{\ket{2^{u}}}_{i+1}\right], 
\\
\hspace{-7mm} {\cal W}^{colored}_i &  = & \sum_{s, t, u =1}^2 \left[P^{\ket{1^{s}}}_{i-1}\,\frac12\left(1^{2^{t}3^{t}}_i-X^{2^{t}3^{t}}_i\right)P^{\ket{1^{u}}}_{i+1} 
                +  P^{\ket{3^{s}}}_{i-1}\,\frac12\left(1^{1^{t}2^{t}}_i-X^{1^{t}2^{t}}_i\right)P^{\ket{3^{u}}}_{i+1}\right], 
\nn \\
\eea  
which consists of the bulk Hamiltonian 
\be
H_{bulk, \,connected}= \sum_{i=1}^n {\cal C}_i + \sum_{i=1}^{n-1}\left[{\cal U}_i + {\cal D}_i + {\cal W}_i^{colored} + {\cal F}_i^{colored}\right].
\label{S32bulkc}
\ee       

The boundary terms (\ref{S32left}) and (\ref{S32right}) are written as 
\begin{eqnarray}
H_{left, \,connected}^{colored} & = & \sum_{s, t=1}^2\left[P^{\ket{2^{s}}}_0P^{\ket{1^{t}}}_1 + P^{\ket{3^{s}}}_0P^{\ket{1^{t}}}_1 + P^{\ket{3^{s}}}_0P^{\ket{2^{t}}}_1\right], 
\label{S32leftc}\\ 
H_{right, \,connected}^{colored} & = & \sum_{s, t=1}^2\left[P^{\ket{1^{s}}}_{n-1}P^{\ket{2^{t}}}_{n} + P^{\ket{1^{s}}}_{n-1}P^{\ket{3^{t}}}_{n} + P^{\ket{2^{s}}}_{n-1}P^{\ket{3^{t}}}_{n}\right].
\label{S32rightc}
\end{eqnarray}
$H_{connected}$ is given by the sum of (\ref{S32bulkc}), (\ref{S32leftc}) and (\ref{S32rightc}).

\subsection{$H_{connected}$ in the second case}
Following the same argument as in the first case, we find that the Hamiltonian is expressed by the sum of 
\be
H_{bulk, \,connected}= \mu\sum_{i=1}^n {\cal C}_i + \sum_{i=1}^{n-1}\left[{\cal U}_i + {\cal D}_i + {\cal F}_i^{balanced}+ {\cal W}_i^{balanced} +{\cal R}^{balanced}\right]
\label{S32bulkc2}
\ee 
and the boundary terms (\ref{S32leftc}) and (\ref{S32rightc}). 
In (\ref{S32bulkc2}), 
\bea
\hspace{-3mm} {\cal F}_i^{balanced} & = & \sum_{s, t =1}^2 \left[P^{\ket{1^s}}_{i-1}\,\frac13\left(1^{1^t2^t}_i-X^{1^t2^t}_i\right)P^{\ket{1^t}}_{i+1} 
+  P^{\ket{1^{s}}}_{i-1}\,\frac13\left(1^{1^t3^t}_i-X^{1^t3^t}_i\right)P^{\ket{1^t}}_{i+1} \right. \nonumber \\
              & &\hspace{7mm}\left.  - P^{\ket{1^s}}_{i-1}\,\frac16\left(1^{2^t3^t}_i-X^{2^t3^t}_i\right)P^{\ket{2^t}}_{i+1}  
               +  P^{\ket{2^s}}_{i-1}\,\frac12\left(1^{2^t3^t}_i-X^{2^t3^t}_i\right)P^{\ket{2^t}}_{i+1}\right], 
\label{S32Fbalanced2}
\\
\hspace{-3mm} {\cal W}_i^{balanced} & = & \sum_{s,t=1}^2 P_{i-1}^{\ket{1^s}}\,\frac12\left(1_i^{2^t3^t}-X_i^{2^t3^t}\right)P_{i+1}^{\ket{1^t}} 
 + \sum_{s,t,u=1}^2 P_{i-1}^{\ket{3^s}}\,\frac12\left(1_i^{1^t2^t}-X_i^{1^t2^t}\right)P_{i+1}^{\ket{3^u}} , \nn \\
\label{S32Wbalanced2}
\\
\hspace{-3mm} {\cal R}_i^{balanced} & = &  \sum_{s=1}^2\sum_{a,b,c=1;\,b>a,c}^3 \left[
P_{i-1}^{\ket{a^s}} P_i^{\ket{b^1}} P_{i+1}^{\ket{c^2}} + P_{i-1}^{\ket{a^s}} P_i^{\ket{b^2}} P_{i+1}^{\ket{c^1}}\right], 
\label{S32Rbalanced2}
\eea
and the rest are the same as in the first case.


\newpage
\begin{center}

\vskip-1.5cm
{\large {\bf Addendum: Area Law Violations and Quantum Phase Transitions in Modified Motzkin Walk Spin Chains} }
\vskip 0.75cm

 Fumihiko Sugino and Pramod Padmanabhan



\vskip 0.5cm 

\end{center}




\noindent
{\bf Keywords:} entanglement entropies, quantum phase transitions, spin chains, ladders and planes, rigorous results in statistical mechanics

\section*{Addendum 1}

We find that the Hamitonian (3.8) contains additional ground states that do not correspond to the SMWs. 
For example, $x_{1,3} x_{3,2} x_{2,1}x_{1,2} x_{2,3} x_{3,1}$ for length 6 enters the negative region but begins and ends on the $x$-axis, whereas the length 6 path $x_{1,3} x_{3,2} x_{2,1}x_{1,3} x_{3,2} x_{2,1}$ goes below the $x$-axis and ends at a negative height. Finally the length 5 path $x_{2,3} x_{3,1} x_{1,2}x_{2,3} x_{3,1}$ stays in the positive quadrant but ends at a positive height above the $x$-axis. These states are valid for the case when $\lambda =0$. 
To lift such ground states, we modify (3.5) and (3.6) as 
\bea
H_{left}  & = & P^{\ket{\left(x_{2,1}\right)_1}} + P^{\ket{\left(x_{3,1}\right)_1}} + P^{ \ket{\left(x_{3,2}\right)_1}} +P^{\ket{\left(x_{1,3}\right)_1,\,\left(x_{3,2}\right)_2,\,\left(x_{2,1}\right)_3}}, \nn\\
H_{right}& = & P^{\ket{\left(x_{1,2}\right)_n}}  + P^{\ket{\left(x_{1,3}\right)_n}} + P^{\ket{\left(x_{2,3}\right)_n}} +P^{\ket{\left(x_{1,2}\right)_{n-2},\,\left(x_{2,3}\right)_{n-1},\,\left(x_{3,1}\right)_n}}. \nn
\eea
Correspondingly, (3.16) and (3.17) become 
\begin{eqnarray}
H_{left, \,connected} & = & P^{\ket{2}}_0P^{\ket{1}}_1 + P^{\ket{3}}_0P^{\ket{1}}_1 + P^{\ket{3}}_0P^{\ket{2}}_1+ P^{\ket{1}}_0P^{\ket{3}}_1P^{\ket{2}}_2P^{\ket{1}}_3, \nn\\ 
H_{right, \,connected} & = & P^{\ket{1}}_{n-1}P^{\ket{2}}_{n} + P^{\ket{1}}_{n-1}P^{\ket{3}}_{n} + P^{\ket{2}}_{n-1}P^{\ket{3}}_{n} + P^{\ket{1}}_{n-3}P^{\ket{2}}_{n-2}P^{\ket{3}}_{n-1}P^{\ket{1}}_n. \nn
\end{eqnarray}
In the $\cS^3_2$ case, (4.14) and (4.15) are modified as 
\begin{eqnarray}
H_{left}  & = & \sum_{s=1}^2~\left[P^{\ket{\left(x^s_{2,1}\right)_1}} + P^{\ket{\left(x^s_{3,1}\right)_1}} + P^{ \ket{\left(x^s_{3,2}\right)_1}}\right] 
+ \sum_{s,s',s''=1}^2 P^{\ket{\left(x_{1,3}^s\right)_1,\,\left(x_{3,2}^{s'}\right)_2,\,\left(x_{2,1}^{s''}\right)_3}}, 
\nn\\
H_{right} & = & \sum_{s=1}^2~\left[P^{\ket{\left(x^s_{1,2}\right)_n}}  + P^{\ket{\left(x^s_{1,3}\right)_n}} + P^{\ket{\left(x^s_{2,3}\right)_n}}\right]
+ \sum_{s,s',s''=1}^2 P^{\ket{\left(x_{1,2}^s\right)_{n-2},\,\left(x_{2,3}^{s'}\right)_{n-1},\,\left(x_{3,1}^{s''}\right)_n}},
\nn
\end{eqnarray}
and (C.7) and (C.8) are 
\begin{eqnarray}
H_{left, \,connected}^{colored} & = & \sum_{s, t=1}^2\left[P^{\ket{2^{s}}}_0P^{\ket{1^{t}}}_1 + P^{\ket{3^{s}}}_0P^{\ket{1^{t}}}_1 + P^{\ket{3^{s}}}_0P^{\ket{2^{t}}}_1\right] 
+\sum_{s,t,u,v=1}^2 P^{\ket{1^s}}_0P^{\ket{3^t}}_1P^{\ket{2^u}}_2P^{\ket{1^v}}_3, 
\nn\\
H_{right, \,connected}^{colored} & = & \sum_{s, t=1}^2\left[P^{\ket{1^{s}}}_{n-1}P^{\ket{2^{t}}}_{n} + P^{\ket{1^{s}}}_{n-1}P^{\ket{3^{t}}}_{n} + P^{\ket{2^{s}}}_{n-1}P^{\ket{3^{t}}}_{n}\right]
+\sum_{s,t,u,v=1}^2 P^{\ket{1^s}}_{n-3}P^{\ket{2^t}}_{n-2}P^{\ket{3^u}}_{n-1}P^{\ket{1^v}}_n. 
\nn
\end{eqnarray}

\section*{Addendum 2}

The Hamiltonian (3.8) changes discontinuously as $\lambda$ increases from zero due to the function $\textrm{sgn} (\lambda)$. 
Here, we can consider to introduce two nonnegative parameters $\lambda_1$ and $\lambda_2$ rather than a single $\lambda$ to make the Hamiltonian changed smoothly. 
First, (3.4) and (3.8) are changed to 
\be
W_{j, j+1} = P^{\frac{1}{\sqrt{2}}\left[\ket{\left(x_{1,2}\right)_j, \left(x_{2,1}\right)_{j+1}} - \ket{\left(x_{1,3}\right)_j, \left(x_{3,1}\right)_{j+1}}\right]} 
+   \lambda_1 \,P^{\frac{1}{\sqrt{2}}\left[\ket{\left(x_{3,1}\right)_j, \left(x_{1,3}\right)_{j+1}} - \ket{\left(x_{3,2}\right)_j, \left(x_{2,3}\right)_{j+1}}\right]}
\nn
\ee
and 
\begin{equation} 
H = H_{left} + H_{bulk} +  H_{right} + \lambda_2\sum_{j=1}^{n-1}~B_{j, j+1},
\nn
\end{equation}
respectively. 
We consider the parameter space consisting of I) $\lambda_1>0$ and $\lambda_2=0$, II) $\lambda_1=\lambda_2=0$ and III) $\lambda_1=0$ and $\lambda_2>0$ 
that preserves the frustration-free property of the Hamiltonian. 
Correspondingly, (3.14) becomes 
\be
 {\cal W}_i = P^{\ket{1}}_{i-1}~\frac12\left(1^{23}_i-X^{23}_i\right)~P^{\ket{1}}_{i+1} 
        +  \lambda_1\,P^{\ket{3}}_{i-1}~\frac12\left(1^{12}_i-X^{12}_i\right)~P^{\ket{3}}_{i+1}. 
        \nn
\ee
The cases I) and III) yield the same ground states as the cases $\lambda=0$ and $\lambda>0$, respectively. 
Therefore, the results of the entanglement entropies computed for those cases are valid without any change. 
For II), we can see that the GSD grows with the length of the chain, which is distinct from the structure of the ground states in I) and III). 
We conclude that the parameter space falls into three phases I), II) and III) as in the figure below, and quantum phase transition occurs at each of the phase boundaries. 

\begin{figure}[h!]
\captionsetup{width=0.8\textwidth}
\begin{center}
		\includegraphics[scale=0.8]{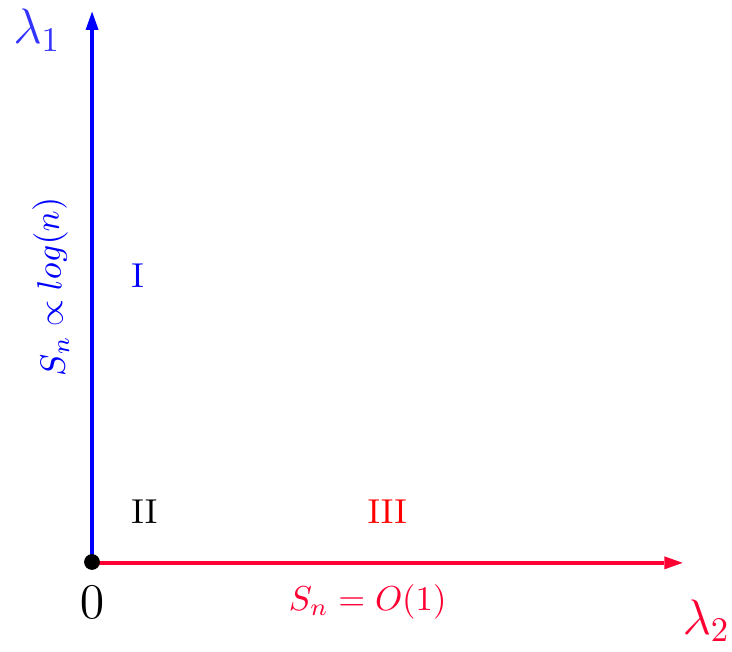} 
	\caption{\small The phases of I, II and III with the change of the entanglement entropy. }
\label{phase}
\end{center}
\end{figure}

\end{document}